\documentclass[fleqn,usenatbib,useAMS]{aa}

\DeclareRobustCommand{\VAN}[3]{#2}
\let\VANthebibliography\thebibliography
\def\thebibliography{\DeclareRobustCommand{\VAN}[3]{##3}\VANthebibliography}

\usepackage{graphicx}	
\usepackage{amsmath}	
\usepackage{amssymb}	
\usepackage[T1]{fontenc}
\usepackage{url,microtype}
\usepackage[switch]{lineno}
\usepackage{aas_macros}
\usepackage{txfonts}
\usepackage{tikz,xcolor}
\usepackage{hyperref}
\hypersetup{
        colorlinks = true,
        urlcolor = blue,
        linkcolor = blue,
        citecolor=blue
}

\newcommand{\ee}{e$^{-}$e$^{+}$}
\newcommand{\ep}{e$^{-}$p$^{+}$}

\graphicspath{{./}{Figures/}}

\DeclareRobustCommand{\orcidicon}{%
	\begin{tikzpicture}
	\draw[lime, fill=lime] (0,0) 
	circle [radius=0.16] 
	node[white] {{\fontfamily{qag}\selectfont \tiny ID}};
	\draw[white, fill=white] (-0.0625,0.095) 
	circle [radius=0.007];
	\end{tikzpicture}
	\hspace{-2mm}
}

\foreach \x in {A, ..., Z}{%
\expandafter\xdef\csname orcid\x\endcsname{\noexpand\href{https://orcid.org/\csname orcidauthor\x\endcsname}{\noexpand\orcidicon}}
}

\begin{document}

    \title{Kinetic simulations of electron-positron induced streaming instability in the context of gamma-ray halos around pulsars}
    
    \titlerunning{Instability of streaming pairs in PWNe}

    \author{Illya Plotnikov\orcidA{}
    \inst{1}
    \and
    Allard Jan van Marle\orcidB{}
    \inst{2}
    \and
    Claire Gu\'epin\orcidC{}
    \inst{2}
    \and
    Alexandre Marcowith\orcidD{}
    \inst{2}
    \and
    Pierrick Martin\orcidE{}
    \inst{1}
    }

   \institute{IRAP, CNRS, OMP, Universit\'e de Toulouse III – Paul Sabatier, Toulouse, France \\
    \email{illya.plotnikov@irap.omp.eu}
   \and
   Laboratoire Univers et Particules de Montpellier (LUPM), Universit\'e de Montpellier, CNRS/IN2P3, CC72, Place Eug\`ene Bataillon, F-34095 Montpellier Cedex 5, France
   }
             
    \date{Received Month XX, 2024; accepted Month XX, 2024}

\abstract{
The possibility of slow diffusion regions as the origin for extended TeV emission halos around some pulsars (such as PSR J0633+1746 and PSR B0656+14) challenges the standard scaling of the electron diffusion coefficient in the interstellar medium.}
{Self-generated turbulence by electron-positron pairs streaming out of the pulsar wind nebula was proposed as a possible mechanism to produce the enhanced turbulence required to explain the morphology and brightness of these TeV halos.}
{We perform fully kinetic 1D3V particle-in-cell simulations of this instability, considering the case where streaming electrons and positrons have the same density. This implies purely resonant instability as the beam does not carry any current.}
{We compare the linear phase of the instability with analytical theory and find very reasonable agreement. The non-linear phase of the instability is also studied, which reveals that the intensity of saturated waves is consistent with a momentum exchange criterion between a decelerating beam and growing magnetic waves. With the adopted parameters, the instability-driven wavemodes cover both the Alfv\'enic (fluid) and kinetic scales. The spectrum of the produced waves is non-symmetric, with left-handed circular polarisation waves being strongly damped when entering the ion-cyclotron branch, while right-handed waves are suppressed at smaller wavelength when entering the Whistler branch. The low-wavenumber part of the spectrum remains symmetric when in the Alfv\'enic branch. As a result, positrons behave dynamically differently compared to electrons. The final drift velocity of positrons can maintain a larger value than the ambient Alfv\'en speed $V_A$ while the drift of electrons can drop below $V_A$. We also observed a second harmonic plasma emission in the wave spectrum. 
An MHD-PIC approach is warranted to probe hotter beams and investigate the Alfv\'en branch physics. We provide a few such test simulations to support this assertion.}
{This work confirms that the self-confinement scenario develops essentially according to analytical expectations, but some of the adopted approximations (like the distribution of non-thermal particles in the beam) need to be revised and other complementary numerical techniques should be used to get closer to more realistic configuration.}


\maketitle

\section{Introduction}
The recent detection of gamma-ray halos around PSR J0633+1746 and PSR B0656+14 (the Geminga pulsar and the pulsar in the Monogem ring) by the HAWC collaboration \citep{HAWC17a, HAWC17b} demonstrates that the propagation of high-energy cosmic rays around their sources can be strongly modified, involving a reduced diffusivity of the particles by more than two orders of magnitude with respect to what is expected from direct measurements of secondary to primary cosmic-ray nuclei ratios. The presence of a gamma-ray halo around Geminga has been since confirmed by Fermi-LAT in the GeV range \citep{DiMauro19}. Again, at these energies, a reduction of the diffusion coefficient by two orders of magnitude seems necessary to explain the halo extension. \citet{DiMauro21} confirm this trend using Fermi data analysis for three supplementary sources from the HAWC collaboration catalogue. More recently, the LHAASO collaboration reported on a similar structure around the middle-aged pulsar J0622+3749 in the $10-100\,{\rm TeV}$ range \citep{LHAASO21}. Finally, the HESS collaboration reported on a detection of a gamma-ray halo around the Geminga pulsar with similar properties deduced from HAWC data \citep{2023A&A...673A.148H}.

As pulsars are anticipated to be sources of cosmic-ray leptons (electrons and positrons), the low particle diffusivity at the origin of gamma-ray halos can have several strong implications. Firstly, the suppression of the diffusion around Geminga and Monogem has called into question the contribution of these sources to the lepton flux detected on Earth \citep{Profumo18, Fang18}. It appears that considering the age of these pulsars (about 340\,kyrs for Geminga and 110\,kyrs for Monogem) the contribution of the confinement zone could provide a local enhancement of the lepton flux which can - at least - explain a fraction of the positron excess observed by the AMS-02 \citep{AMS14} and the PAMELA \citep{PAMELA09} experiments \citep{Fang18, Tang19}. The same trend is found by \cite{DiMauro19} considering a time-dependent release of positrons and accounting for pulsar motion. The authors find that Geminga / Monogem pulsars can at most contribute to $20\%$ / a few \% of the positron flux at high energy detected on Earth. Secondly, if the pulsar gamma-ray halo is a general phenomenon, then the cosmic ray content and gamma-ray flux are expected to be enhanced in the galactic disc following the source distribution \citep{Johannesson19}. \citet{2023PhRvD.107l3020S} invoke the possible effect due to radiation losses to produce softer spectra for electron-positron pairs escaping a gamma-ray halo. The loss-softening effect seems necessary to explain the positron spectrum and fraction observed on Earth \citep{PhysRevD.103.083010}. \cite{Manconi20}, using a two-zone model combined with the ATNF pulsar distribution, can explain AMS-02 lepton data with a reasonable fraction of pulsar spin-down luminosity converted into electron-positron pairs of about $10\%$. However, this possibility is questioned by \citet{Martin22a}. These authors also used a two-zone diffusion model and show that assuming that gamma-ray halos are common around middle-aged pulsars may lead to overshooting the local positron flux measured with the AMS-02 experiment. \citet{Martin22a} estimate that at most a fraction of 5-10\% of nearby middle-aged pulsars could develop a gamma-ray halo. In total, in the Galaxy the upcoming Cherenkov Telescope Array is estimated to be able to detect about 100 gamma-ray halos \citep{Martin22b}. \\

The origin of the slow particle diffusion remains a subject of active debate. Several interpretations of the HAWC observations have been put forward so far. \citet{Recchia21} interpreted the HAWC gamma-ray intensity profiles as a combination of ballistic and diffusive particle motions, in a model that does not invoke any specific reduced diffusion to explain the gamma-ray data. Yet, accounting for the observed level of gamma-ray emission requires a power input that exceeds the available spin-down luminosity in some cases \citep{Bao:2021}. Alternatively, perpendicular diffusion could be the origin of the slow diffusion, and favourable magnetic field line orientation with respect to the line of sight produces the observed halos, but the scenario seems more and more unlikely as we discover a growing number of such objects~\citep{DeLaTorreLuque:2022}. \citet{Lopezcoto18} argue that the reduced diffusivity is due to specific properties of the magnetic turbulence surrounding the pulsars. In particular, the gamma-ray data are well-fitted if the turbulence coherence length is $\sim 1\,{\rm pc}$, hence reduced by one or two orders of magnitude with respect to its expected value in the interstellar medium, as deduced from Faraday rotation and depolarisation data \citep{Haverkorn08}. One possible origin of this reduced coherence length is that the pulsar still finds itself in a medium under the influence of the associated supernova remnant turbulence \citep{2019MNRAS.488.4074F}. Finally, \cite{2018PhRvD..98f3017E} investigate the triggering of the resonant streaming instability by an electron-positron (\ee) beam injected at the pulsar wind termination shock. The authors considered the injection of an \ee\ power-law momentum distribution as $p^{-3.5/-3.2}$ from the pulsar termination shock and solved a system of coupled kinetic equations for cosmic rays and triggered waves including turbulent wave damping. They find a reduced diffusivity within a radius of about $10-20\,{\rm pc}$ around the source for time scales of the order of $100$\,kyr. These results were revised (and corrected) by \citet{Mukho22}, who showed that a reduced diffusion region can be sustained over larger distances and longer duration, encompassing the properties of the Geminga system, even when accounting for ion-neutral wave damping and assuming a flatter particle injection spectrum. 

The latter work reveals that the self-confinement scenario relies on several assumptions for it to be relevant. Essentially, it requires that maximum power is channelled into the pair stream to provide a growth of the turbulence up to a sufficient level. This implies: (i) a beam starting right after the pulsar's birth; (ii) one-dimensional injection along a flux tube; (iii) no proper motion of the pulsar; and (iv) an injection spectrum peaking around 10-100\,TeV, the energy where reduced diffusion is the most solidly constrained. In realistic conditions, it is not guaranteed that these favourable requirements are met: for (i), injection initially occurs in the pulsar wind nebula, and pairs may not be immediately available to the ambient medium until the system has evolved to a bow-shock nebula stage, about several thousands or tens of thousands of years later, at which time the pulsar has lost much of its spin-down power \citep{Gaensler:2006}; for (ii) and (iii), realistic conditions such as three-dimensional expansion and proper motion of the pulsar imply beam dilution, hence less power for exciting the instability; and for (iv), particle spectra produced by pulsars and their nebulae are observationally constrained to peak around 100\,GeV \citep{Torres:2014}. 

Nevertheless, \citet{Mukho22} show that the self-confinement option remains a viable explanation in some conditions, all the more so that it can be combined with other scenarios (pre-existing fluid turbulence imparted by the expansion of the remnant\footnote{Pre-existing turbulence could help because it may provide a supplementary source of particle scattering if acting on an appropriate scale. Yet, this effect has to be assessed with a dedicated study. }, or kinetic turbulence triggered by the acceleration of protons at its forward shock). For that reason, we consider that the underlying physics requires a more detailed understanding.
 
In this work, we investigate the case of electron-positron driven instabilities in a similar framework as proposed by \citet{2018PhRvD..98f3017E} but we extend it to the case of higher electron-positron to background plasma density ratios, thus discarding the test-particle assumption\footnote{In this work, we use the test-particle terminology following \citet{2018PhRvD..98f3017E}. It is valid as long as the electron-positron beam is not modifying the dispersion relation of Alfv\'en waves, that is when the amplitude of excited waves is small: $\delta B \ll B_0$. When this limit is not satisfied, the instability drives either strongly modified Alfv\'en waves or waves or different nature and scale (e.g., kinetic Alfv\'en waves, Whistler, ion-cyclotron, etc.)}. This regime could be applicable to the instability development in Pulsar wind nebula during early evolutionary phase when the density of released pairs is the highest. Our study concentrates on the microphysics of both linear and non-linear phases of the pair-driven streaming instability by means of kinetic particle-in-cell (PIC) numerical simulations. Contrary to a recent study by \citet{Gupta21} we consider only the case where the density of streaming electrons and positrons is equal. Thus, driving only resonant instability and discarding faster-growing non-resonant instability. The main objective of this work is to explore the microphysics of the self-confinement of a pair stream \citep[general review on the streaming instability in astrophysical plasmas can be found in][]{2021PhPl...28h0601M}. Exporting that to a realistic model of a pulsar halo is deferred to future studies.

The layout of this paper is as follows: in section~\ref{sect:LIN} we derive the general expression for the linear growth rate and give estimates in the case of a Maxwell-Jüttner distribution. We also give an estimate of the magnetic field saturation level in the case of the resonant streaming instability. In section~\ref{sect:pic} we present the results of PIC simulations. We further provide some interpretation.
Section~\ref{sect:DIS} proposes a discussion before concluding.

\section{Streaming instability linear growth rates}\label{sect:LIN}

\begin{table}
\caption{Notations used in this work.}
\label{table:notations}
\begin{tabular}{l l}
\hline
\hline
Quantity & Notation \\
\hline
Light speed, elementary charge & $c,\;e$ \\
Frequency, wavenumber & $\omega,\;k$ \\
Magnetic field strength & $B$ \\
Particle species & $\alpha 
\; {\rm (e,i,cr)}$ \\
(electron, proton, CRs)) & \\
Mass, charge, momentum & $m_\alpha,\;q_\alpha,\;p_\alpha$ \\
Lorentz factor & $\gamma_\alpha = \sqrt{1+(p_\alpha/m_\alpha c)^2}$ \\
Gyrofrequency & $\Omega_\alpha = q_\alpha B / \gamma_\alpha m_\alpha c = \Omega_{c \alpha} / \gamma_\alpha$ \\
Gyroradius & $R_{\rm L,\alpha} = c / \Omega_{c \alpha}$ \\
Background density & $n_i = n_e$ \\
Beam density & $n_{\rm cr}$ \\
Alfv\'en velocity & $V_A = B / \sqrt{4 \pi n_i m_i}$ \\
Drift velocity & $V_D$ \\
Electron/ion temperature & $T_{e,i} = \theta_{\rm bg} m_e c^2 / k_B$ \\
(background) & \\
Electron plasma frequency & $\omega_{\rm pe} = \sqrt{4 \pi n_e e^2 / m_e}$ \\
Electron skin-depth & $d_{\rm e} = c / \omega_{\rm pe}$ \\
 \hline
\end{tabular}
\end{table}

\citet{2018PhRvD..98f3017E} consider a particular case of a beam composed of electrons and positrons. Here, we do not first assume any particular composition of the drifting beam. This allows us to present below the general formalism necessary to derive the growth rate of the streaming instability. This physical set-up is different from the case of a mildly-relativistic electron-positron spine injected into a magnetised electron-proton plasma \citep{Dieckmann18, Dieckmann19, 2022PhPl...29i2103D}. A situation where our formalism does not apply, because both plasmas are initially separated by hydrodynamical structures. The formalism developed below could however be of interest, but at later stages if some mixing applies between the two plasma components. Hence, potentially, it could also be applied to compact object jets.

\subsection{General form of the dispersion relation}\label{sec:growth}
We adopted CGS units and all notations used in this work are reported in \autoref{table:notations}. The dispersion relation for modes of frequency $\omega$ and wavenumber $k$ can be written as \citep{2009MNRAS.392.1591A}:
\begin{equation}
    {k^2 c^2 \over \omega^2} = 1 + \chi \ .
\end{equation}
The plasma response function (or susceptibility) $\chi$ has two components, one for the background $\chi_{\rm bg}$ and one for the CR beam $\chi_{\rm cr}$. We conduct the calculation in the beam rest frame in which the CR distribution is assumed to be isotropic. Each beam component is assumed to drift at the same speed $V_D$ with respect to the background plasma assumed to be cold. We first calculate the CR contribution $\chi_{\rm cr}$. Assuming an isotropic distribution in the momentum space $g(p)$ given by $f_{\rm cr} = n_{\rm cr} g(p)/4\pi$ we have 
\begin{eqnarray}\label{Eq:XICR}
    \chi_{\rm cr}&=&  \sum_{\alpha} {\pi q_\alpha^2 \over \omega}  n_{{\rm cr},\alpha} \int_{p_0}^{p_{\rm max}} dp p^2 \upsilon(p) {d g_{\alpha}(p) \over dp}  \nonumber \\
 && \times \int_{-1}^{1} {(1-\mu^2) \over \omega + k\upsilon(p) \mu \pm \Omega_{\rm \alpha}} d\mu\ ,
\end{eqnarray}
where $\mu$ is the particle pitch-angle cosine. We only consider the fastest growing modes propagating parallel to the background large-scale magnetic field, namely we set $k_\parallel = k$ in the above expression. 
Each particle species has a distribution $g_\alpha(p)$ such that $\int dp p^2 g_\alpha(p)= 1$; different forms of particle distribution are discussed below. Finally, $n_{\rm cr, \alpha}$ is the CR density. If the calculation is carried in the background plasma frame, a supplementary term $(1-kV_D/\omega)$ must come in front of Eq. \ref{Eq:XICR}. This term is necessary to quench the instability if the drift speed matches the phase speed of the destabilised modes. 
For a particular species $\alpha$ the integral over $\mu$ can be derived using the Plemelj formula:
\begin{eqnarray}
   \int_{-1}^{1} {(1-\mu^2) \over \omega + k\upsilon \mu \pm \Omega_{\rm \alpha}} d\mu &=& \mathcal{P} \int_{-1}^{1} {(1-\mu^2) \over \omega + k\upsilon \mu \pm \Omega_{\rm \alpha}} d\mu \\
   && -i\pi \int_{-1}^{1} (1-\mu^2) \delta(\omega + k\upsilon \mu \pm \Omega_{\rm \alpha}) d\mu \nonumber \ ,
\end{eqnarray}
where the particle speed is a function of the momentum $p$ and can be written as $\upsilon(p)$.
We find:
\begin{eqnarray}\label{Eq:DRES0}
\chi_{\rm cr} &=& {\pi e^2 \over k \omega}\sum_{\alpha} N_{CR,\alpha} \int_{p_0}^{p_{\rm max}} dp p^2 {d g_{\alpha}(p) \over dp}  \\
&& \left[ \left(1-\mu_{\rm R, \alpha}^2\right) \log\left|{1-\mu_{\rm R, \alpha}^{-1} \over 1+ \mu_{\rm R, \alpha}^{-1}}\right|-2\mu_{\rm R, \alpha} -i\pi \left(1-\mu_{\rm R, \alpha}^2 \right)\right] \nonumber \ ,
\end{eqnarray}
where the wave-particle resonance selects a particular pitch-angle cosine
\begin{equation}
    \mu_{\rm R, \alpha} = \mp {\Omega_{\rm \alpha} \over k\upsilon} -{\omega \over k\upsilon} \, .
\end{equation}
We have at this stage considered that the beam is composed of particles with charges $q = \pm e$. \autoref{Eq:DRES0} has three terms under the p-integral: the two first terms are connected to the CR current and are associated with the non-resonant branch of the streaming instability, the last term is associated with the resonant branch.

The background electron-proton (\ep) plasma contribution is given by Eq. 14 in \citet{2018PhRvD..98f3017E}. Considering that each beam component is cold and drifts at the same speed $V_D$ we have:
\begin{equation}\label{Eq:chibg_1}
    \chi_{\rm bg} =-{4\pi e^2 \over \omega^2} \left({n_i \over m_i}{\omega + k V_D  \over \omega + k V_D \pm \Omega_{\rm ci}} +  {n_e \over m_e}{\omega + k V_D  \over \omega + k V_D \pm \Omega_{\rm ce}}
    \right) \, .
\end{equation}
It is easy to add any ion component (e.g., Helium) to this equation. The background is considered as cold enough that the thermal motion does not induce any modification of the linear growth rate. In the interstellar medium (ISM) this would correspond to warm or cold media. For pulsar propagation into hotter media, like superbubbles, thermal background effects are expected to alter the growth of oblique modes \citep{Foote79}. The non-resonant branch of the instability is also modified in that case \citep{2010ApJ...709.1412Z, 2021MNRAS.500.2302M}.

\subsection{Low frequency limits}\label{sec:growthee}
We now assume that $(\omega + kV_D) \ll \Omega_{\rm ci} \ll \Omega_{\rm ce}$. Hence, if we have $n_e= n_i$, the contribution from the background \ep plasma is
\begin{equation}\label{Eq:chibg}
    \chi_{\rm bg} \simeq {(\omega + k V_D )^2  \over \omega^2}  {c^2 \over V_A^2}.
\end{equation}
For cosmic rays contribution, we need to specify the distribution function $g(p)$. In this work we considered a Maxwell-J\"uttner (MJ) distribution unless otherwise specified. The results for the power-law and mono-energetic cases are reported in appendix \ref{sec:appendixgrowth}. The solutions in the case of a MJ distribution are used in the numerical experiments in section~\ref{sect:pic}. We restrict ourselves to the case of a pure electron-positron beam with the same density for each species. More complex cases involving different electron and positron densities or momenta and the addition of a hadronic component deserve future works.

We assume a beam composed of electrons and positrons with $n_{\rm cr,+}= n_{\rm cr,-} = n_{\rm cr}$ (and $g_+ = g_-= g$) and $\omega \ll \Omega_{\rm e^+}$. The non-resonant term vanishes and we recover the results by \citet{2018PhRvD..98f3017E} (their Eq. 17), namely:
\begin{equation}
    \chi_{\rm cr} = -2 {i \pi^2 e^2 \over k \omega} n_{\rm cr} \int_{p_{\rm min}}^{p_{\rm max}} dp {dg(p) \over dp} (p^2 - p_{\rm min}(k)^2) \ ,
    \label{eq:chi_CRMJ}
\end{equation}
where the maximum particle momentum is $p_{\rm max}$ and 
$p_{\rm res}(k) = {m_e |\Omega_{\rm ce}| \over k}$. The factor 2 in front of the RHS part of Eq. \ref{eq:chi_CRMJ} comes out as we have $n_{\rm cr,+}+n_{\rm cr,-}= 2 n_{\rm cr}$. For simplicity and tractability, we adopt the ultra-relativistic limit $E \simeq pc \gg m_e c^2$. The MJ distribution is then defined as:
\begin{equation}\label{Eq:MJ}
    g(p) = \frac{1}{m_e^3c^3}\frac{1}{\theta K_2({1 \over \theta})} e^{-\gamma(p)/\theta} \, ,
\end{equation}
where $\theta = k_B T_{\rm cr} / (m_e c^2)$, $k_B$ is the Boltzmann constant, $T_{\rm cr}$ is the temperature of CR electrons (same for CR positrons), and $K_2$ is the modified Bessel function of the second kind. The particle Lorentz factor is $\gamma(p)$. The distribution for positrons is the same with a density $n_{\rm cr,+}$. 

The choice of MJ distribution instead of a power-law is motivated by three main aspects. The first one is that MJ is physically motivated and allows to restrict the range of particle Larmor radii, hence to better isolate the resonant modes in our simulations. The second one is that for broad distributions extending to very high energies -- such as power-law distributions -- excessive numerical resources are required with the PIC method to cover sufficient dynamical scales in energy or momentum. The third reason is that MJ distribution is analytically tractable. This allows direct comparisons of the growth rate with numerical results (control of spurious non-physical effects).

Evaluating the integral in Equation \ref{eq:chi_CRMJ} we find 
\begin{eqnarray}
    \chi_{\rm cr} &\simeq& i\frac{4\pi^2 e^2}{\omega k} \frac{n_{\rm cr}}{m_ec} \frac{\theta}{K_2({1 / \theta})} I(k,\gamma_{\rm min}) \ , \nonumber \\
    I(k,\gamma_{\rm min}(k)) &=& \exp(-{\gamma_{\rm min}\over \theta}) \left({\gamma_{\rm min}(k) \over \theta}+1\right), \nonumber \\
    \gamma_{\rm min}(k) &=& \sqrt{1+\left({\Omega_{\rm ce}\over kc}\right)^2} \ , 
\end{eqnarray}
and $\gamma_{\rm min}= \gamma(p_{\rm min})$. We have assumed that $\gamma(p_{\rm max}) \gg \theta$.\\
The dispersion relation then reads as
\begin{equation}
k^2 V_A^2 = \omega^2 \left({V_A \over c}\right)^2+\left(\omega+kV_D\right)^2 + i \pi  {\omega \, \Omega_{\rm ci} \over k \, R_{\rm L,e}} {n_{\rm cr} \over n_i} {\theta \over K_2({1 / \theta})} I(k,\gamma_{\rm min})  \ ,
\end{equation}
where $R_{\rm L,e}=c/\Omega_{\rm ce}= R_{\rm L,cr}/\theta$ and $R_{\rm L,cr} = k_B T_{\rm cr}/e B$. The instability growth rate is given by the system 
\begin{equation}
 k^2 V_A^2 \simeq \tilde{\omega}^2 + i  \omega {n_{\rm cr} \over n_i} \Omega_{\rm ci} I_2(k) \ , 
\end{equation}
with $\tilde{\omega}= \omega + kV_D$, $I_2(k, \theta)= {\pi \over k R_{\rm L,e}} {\theta \over K_2(1/\theta)} e^{-{\gamma_{\rm min}\over \theta}} \left(1+\frac{\gamma_{\rm min}}{\theta}\right)$. The system is of second order as we assume $\Re (\tilde{\omega}) \ll kV_D$ \citep{2018PhRvD..98f3017E}. In the test-particle limit (TPL) $\sigma I_2(k,\theta) \ll k^2 V_A^2$, $\sigma = kv_D {n_{\rm cr} \over n_i} \Omega_{\rm ci}$, the linear growth rate $\Gamma \equiv \Im (\tilde{\omega})$ reads as
\begin{equation}
\Gamma \simeq {\pi \over 2} \Omega_{\rm ci}{1 \over k R_{\rm L,cr}} {V_D \over V_A} {n_{\rm cr} \over n_i} {\theta^2 \over K_2(1/\theta)} e^{-{\gamma_{\rm min}\over \theta}} \left(1+\frac{\gamma_{\rm min}}{\theta}\right)  \ .\label{Eq:growth_rate_final}
\end{equation}
and the dispersion relation is almost unmodified $\Re (\tilde{\omega}) = k V_a$. In the relativistic regime $\gamma_{\rm min}(k) \simeq \theta/(kR_{\rm L,cr})$ and $\theta^2/K_2(1/\theta) \rightarrow 0.5$ (for $\theta \gg 1$). We note that the dispersion in \autoref{Eq:growth_rate_final} can be derived using alternative approaches as it is the case in \citet{Bai19, Plotnikov21}.

\subsection{Higher frequency solutions: polarisation-dependent growth rates}
As the simulations presented below depart from the TPL and have limited dynamical range in wavenumber, it is useful to relax the assumption $(\omega + kV_D) \ll \Omega_{\rm ci}$. In this case, processes at frequencies $\Omega_{\rm ci} < \omega < \Omega_{\rm ce}$ are now included (e.g., ion-cyclotron resonances, excitation of Whistler waves). Yet, the two following approximations remain valid: $\Omega_{\rm ci} \ll \Omega_{\rm ce}$  and $\omega \ll \omega_{\rm pe}$. The former is always satisfied for $m_i \gg m_e$ and the latter avoids looking for the resonances on the light-wave branch. In this case, the general contribution from the background plasma
\begin{equation}
    \chi_{\rm bg} = \frac{c^2}{V_A^2} \frac{\tilde{\omega}^2}{\omega^2} \frac{ 1- \Omega_{\rm ci}/\Omega_{\rm ce}}{[1 \pm \tilde{\omega}/\Omega_{\rm ci}][1 \pm \tilde{\omega}/\Omega_{\rm ce}]}\, ,
\end{equation}
leads to a third-order dispersion relation
\begin{equation}
    k^2 V_A^2 = \omega^2 {V_A^2 \over c^2} + \frac{\tilde{\omega}^2}{[1 \pm \tilde{\omega}/\Omega_{\rm ci}][1 \pm \tilde{\omega}/\Omega_{\rm ce}]} + \frac{i \omega \mathcal{C}}{k V_D}  \ ,
\end{equation}
where $\mathcal{C} \equiv \pi V_D  {\Omega_{\rm ci} \over R_{\rm L,cr}} {n_{\rm cr} \over n_p} {\theta^2 \over K_2({1\over \theta})} I(k,\gamma_{\rm min})$. After reorganising the terms
\begin{eqnarray}\label{Eq:HFDR}
     && \tilde{\omega}^3 \frac{i \mathcal{C}}{k V_D} + \tilde{\omega}^2 \left(\Omega_{\rm ce} \Omega_{\rm ci} - (k V_A)^2 - i\mathcal{C} \pm \frac{i \mathcal{C} \Omega_{\rm ce}}{k V_D}\right) \nonumber \\
     && +\,\tilde{\omega} \, \Omega_{\rm ce} \left( \mp (k V_A)^2 \mp i\mathcal{C} + \frac{i \mathcal{C} \Omega_{\rm ci}}{k V_D}  \right) \nonumber \\
     && -\,\Omega_{\rm ce} \Omega_{\rm ci} \left[(k V_A)^2 + i\mathcal{C}\right] = 0 \ .
\end{eqnarray}
Solutions of this equation can be calculated by considering the intersection of the real and imaginary planes solutions.

\subsection{Magnetic field saturation}
The resonant streaming instability has different saturation criteria. \citet{Holcomb19} discuss the saturation process in some details. The saturation is a two-step process. First, the linear growth rate at the fastest-growing wavenumber saturates. Then, in a second time, the drop of the growth over all wavenumbers induces a reduction of the mean drift speed of the CR distribution down to the local Alfv\'en speed. This second step properly corresponds to the instability saturation. In order to evaluate the total magnetic field strength at saturation we adopt the criterion proposed by \citet{Holcomb19, Bai19} which balances the total momentum density lost by CRs  $\Delta P_{\rm cr} \simeq (4 / 3) [(V_D-V_A) / c] n_{\rm cr} \langle p \rangle$ with the momentum density acquired by forward propagating Alfv\'en waves $\Delta P_{\rm w} \simeq \rho V_A (\delta B^2 / B_0^2) $. Here, $n_{\rm cr}$ is the CR density and $\langle p \rangle$ the mean CR momentum, $\rho$ is the local gas (thermal proton) density. We find:
\begin{equation}\label{Eq:BSAT}
    {\delta B^2_{\rm sat} \over B_0^2} \simeq {8 \over 3} {(V_D-V_A) \over V_A} {n_{\rm cr} \over n_i} {m_e \over m_i} {\langle \gamma v \rangle \over c} \, .
\end{equation}
The above criterion likely provides an upper limit to the perturbed magnetic energy as it assumes that all momentum from CR anisotropy is transferred to Alfv\'en waves. Some amount of momentum is transferred to the background gas (mean momentum and heating), forming the background fluid `wind' \citep{Holcomb19}. Besides, numerical simulations always have a minimum of dissipation.
                                         
\section{Particle-in-cell simulations}\label{sect:pic}

\subsection{Simulation method and setup}

In this section, we use the PIC code \textsc{Smilei} \citep[][]{Derouillat18} to perform 1D3V simulations, where we retain one spatial direction, but all three components of velocities and electromagnetic fields (1D in space and 3D in momentum). The use of 1D geometry is justified by the expectation that the most dominant wavemodes driven by the instability are parallel to the external magnetic field (slab-type Alfv\'en modes with $\mathbf{k} \parallel \mathbf{B_0} $, see Sect. \ref{sec:growth}).\\

We set the simulation in the following way. The 1D box is filled with background electron-ion cold plasma at rest and a streaming beam of hot electron-positron plasma. The electron-positron pairs are inserted randomly in the simulation box. The number density ratio of the beam to background is set to be small, $n_{\rm cr}/n_i \ll 1$. The external magnetic field is parallel to the simulation box $\mathbf{B} = B_0 \mathbf{e}_x$. The typical setup and simulation parameters are listed as follows:

\begin{itemize}
\item The drifting beam (CRs) is composed of pairs and the background static plasma is composed of electrons and ions. Both components are charge and current neutral.
\item We employ periodic boundary conditions for fields and particles.
\item The box size is $L_x = 11264 c / \omega_{\rm pe} = 11264 d_e$. This ensures that the grid length corresponds to at least a dozen of fastest-growing wavemodes.
\item  The electron skin depth, $d_e = c /\omega_{\rm pe}$, is resolved with 6 grid points. The time step is $\Delta t = 0.95 \Delta x / c$. (corresponding to CFL number equal to 0.95). We also tested twice smaller time step with no observed difference in simulation results.
\item There are 100 macro-particles per cell per species, and we use 4-th order macro-particle shape factors \footnote{In this case, the charge and current of one macro-particle is distributed over 5 spatial grid-points \citet{Derouillat18}.}. This ensures a reasonably low level of Poissonian noise. As the physical density of the beam pairs is much smaller than the density of the background (i.e., $n_{\rm cr} \ n_i \ll 1$) we use different macro-particle weights for different species. This effectively decreases numerical noise and increases the statistical significance of the distribution of beam particles \citep[see][]{Holcomb19}.
\item The typical mass ratio  $m_i / m_e = 100$. But we also explored 1, 200, 300, 400, and 1000.
\item The background plasma temperature $T_e = T_i = 10^{-3} m_e c^2$, that is $\theta_{\rm bg} = 10^{-3}$ (in order to get the cell size not too far from the Debye length). The temperature is the same for ions and electrons.
\item Number density ratio of CR (positrons) to background ions $n_{\rm cr}/n_i$ between $10^{-3}$ and $10^{-2}$. As the beam and background plasma are both neutral, this implies that the ratio of CR electrons to background electrons ($n_{\rm cr}/n_e$) is the same. The reference run -- also called Fid run -- corresponds to $n_{\rm cr}/n_i = 10^{-2}$. 
\item The initial distribution function of the beam is drifting Maxwell-J\"uttner distribution (see Eq. \ref{Eq:MJ}) with temperature of $T_{\rm cr}$ (typically $\theta = 10$) and mean speed in $x$-direction is $V_D = 0.1c$. This sets the typical Larmor radius $R_{\rm L,cr} $ of beam electrons (ans positrons).
\end{itemize}

The simulation can be evolved up to $t\omega_{\rm pe} = 10^6$, hat generally corresponds to $t \Omega_{\rm ci} = 10^3$ (for the case of Fid run: $m_i/m_e = 100$ and $\omega_{\rm pe} / \Omega_{\rm ce} = 10$). In some cases, limited computational resources imposed the simulation to stop earlier. In the \autoref{table:PICparam} we provide the typical parameters for different simulations presented in this study.

We note that in the Fid run, plasma beta parameter is $\beta \simeq 8\pi n_i k_B T_i / B_0^2 = (\upsilon_{\rm th,i} / V_A)^2 = 0.2$. This value remains unchanged for different mass ratios $m_i / m_e$ and different $n_{\rm cr} / n_i$, explored in PIC simulations here. Hence, $\beta = 0.2$ is maintained for simulations PIC-1 to PIC-15. 

\begin{table}
      \caption{PIC simulation parameters presented in this work. The fiducial (or Fid) run is PIC-1.}
      \label{table:PICparam}
    \begin{tabular}{ c|c c c c c c c }
    \hline
     \hline
        Run ID & $ {n_{\rm cr} \over n_i}$ & ${m_i \over m_e}$ &  ${\Omega_{\rm ce} \over \omega_{\rm pe}}$ & ${V_D \over V_A}$ & $\theta_{\rm cr}$ & ${t_{\rm sim} \omega_{\rm pe}}$ \\ 
        \hline
        PIC-1 & 0.01 & 100 &  0.1 & 10 & 10 & $10^6$ \\
        {\bf (Fid)} & &  &   &  &   &   &  \\
        PIC-2 & 0.007 & 100 & 0.1  & 10 & 10 & $3 \times 10^5$ \\ 
        PIC-3 & 0.005 & 100 &  0.1 & 10 & 10  & $5 \times 10^5$  \\ 
        PIC-4 & 0.004 & 100 &  0.1 & 10 & 10  & $5 \times 10^5$  \\
        PIC-5 & 0.003 & 100 &  0.1 & 10 & 10  & $10^6$   \\ 
        PIC-6 & 0.002 & 100 &  0.1 & 10 & 10   & $10^6$  \\ 
        PIC-7 & 0.001 & 100 &  0.1 & 10 & 10   & $10^6$  \\
        \hline
        PIC-8 & 0.01 & 200 &  0.1 & 14.1 & 10  & $3\times 10^5$  \\
        PIC-9 & 0.01 & 300 &  0.1 & 17.3 & 10 & $10^6$ \\
        PIC-10 & 0.01 & 400 &  0.1 & 20 & 10   & $6 \times 10^5$ \\
        PIC-11 & 0.01 & 1000 &  0.1 & 31.6 & 10   & $10^6$ \\
        \hline
        PIC-12 & 0.005 & 50 &  0.1 & 7.1 & 10   & $3\times 10^5$ \\
        PIC-13 & 0.005 & 200 &  0.1 & 14.1 & 10   & $10^6$  \\ 
        PIC-14 & 0.005 & 300 &  0.1 & 14.1 & 10   & $10^6$  \\ 
        PIC-15 & 0.005 & 1000 &  0.1 & 31.6 & 10    & $10^6$ \\
         \hline
        PIC-16 & 0.01 & 100 &  0.05 & 20 & 10    & $10^6$ \\
         \hline
        PIC-17 & 0.001 & 1 &  0.01 & 14.1 & 1  & $10^6$   \\
         {\bf ($e^{\pm}$ bg)} & &  &   &  &   &   &  \\ 
         \hline
         \hline
    \end{tabular}
\end{table}

 \subsection{Simulation results}
 In this section, we describe the results of PIC simulations of the instability. We start with a description of the overall time-evolution of the Fid simulation, and then show how the linear growth rates compare to the analytical predictions. We then discuss the saturation of the instability and some properties of the self-generated wave spectrum.  

  \subsubsection{Overall evolution}
  \label{subsect:Overall_PIC}

  \begin{figure}
	\begin{center}
	\includegraphics[width=0.95\columnwidth]{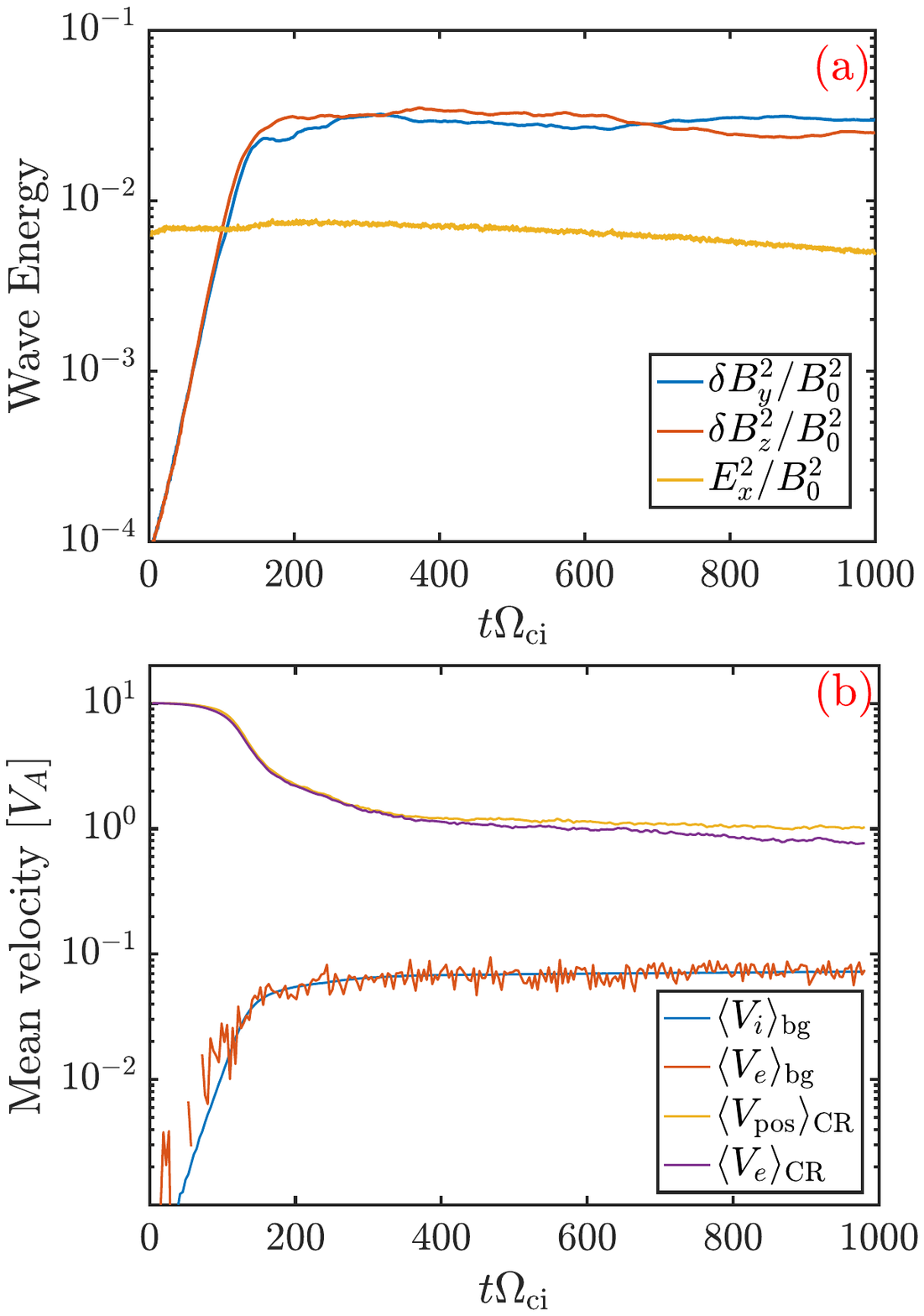}
	\caption{Overall evolution of the instability in time, for the Fid run. {\emph{Top panel:}} wave energy evolution. Blue and red lines depict the evolution of transversal perturbations $\delta B_y^2 / B_0^2$ and $\delta B_z^2 / B_0^2$, respectively. The yellow line corresponds to the level of electrostatic fluctuations $E_x^2 / B_0^2$. {\emph{Bottom panel:}} mean velocity evolution of different species. Blue and red lines correspond to the background ions and electrons, respectively. The yellow and magenta lines show the evolution of the drift velocity of beam positrons and electrons, respectively.} 
	\label{fig:overall_evolution_Fid}
	\end{center}
\end{figure}
  
  The simulation starts with an anisotropic distribution of CRs. This anisotropy is small ($\delta f / f_0 \ll 1$) and is equivalent to an average drift velocity in $x$-direction that is small compared to $c$, 
  $V_D \ll c$. If the initial drift is super-Alfv\'enic, $V_D > V_A$, then the instability is triggered. As a result, magnetic field grows exponentially during the so-called linear phase of the instability. At some point this growth ends (start of the post-linear phase) and, eventually, the level of magnetic fluctuations reaches a maximum (start of the saturated phase). 

In \autoref{fig:overall_evolution_Fid}, we show the evolution of the magnetic wave energy $\delta B^2/B_0^2$ (top panel) and the bulk velocities of different species of the beam-plasma system (bottom panel). This figure shows the result of the Fid simulation.
The linear phase of the instability is clearly identifiable between $t\Omega_{\rm ci} = 10$ and $100$, where the wave energy growth is exponential (the scale of the $y$-axis is logarithmic). Immediately after the linear phase, the amplitude of magnetic fluctuations becomes large enough to scatter beam electrons and positrons efficiently. As a result, the streaming velocity of beam species decreases rapidly from $V_D = 10V_A$ to $V_D \simeq 2 V_A$ (see orange and magenta lines in the bottom panel). At some point (around $t\Omega_{\rm ci} = 150$) the wave growth ends, and the instability enters the saturated phase.
During the saturated phase (at $t\Omega_{\rm ci} > 150$) the evolution becomes slow as the isotropization process of remaining particles takes more time. At the final time of the simulation ($t\Omega_{\rm ci} = 10^3$) the level of magnetic fluctuations is $\delta B^2 / B_0^2 = (\delta B_y^2 + \delta B_z^2) / B_0^2 \simeq 0.06$, while the drift velocity of beam positrons decreased to $\sim V_A$ (i.e., isotropic in the wave-frame). The velocity of beam electrons is slightly below $V_A$. We impute this effect to the observation that some part of self-generated waves in non-Alv\'enic (see detailed analysis of wave properties later in the article). We note that some momentum was also transferred to the background fluid, both electrons and ions reach $\langle V_i \rangle_{\rm bg} \simeq \langle V_e \rangle_{\rm bg} \simeq 0.07 V_A$.

\begin{figure*}
	\begin{center}
	\includegraphics[width=0.99\textwidth]{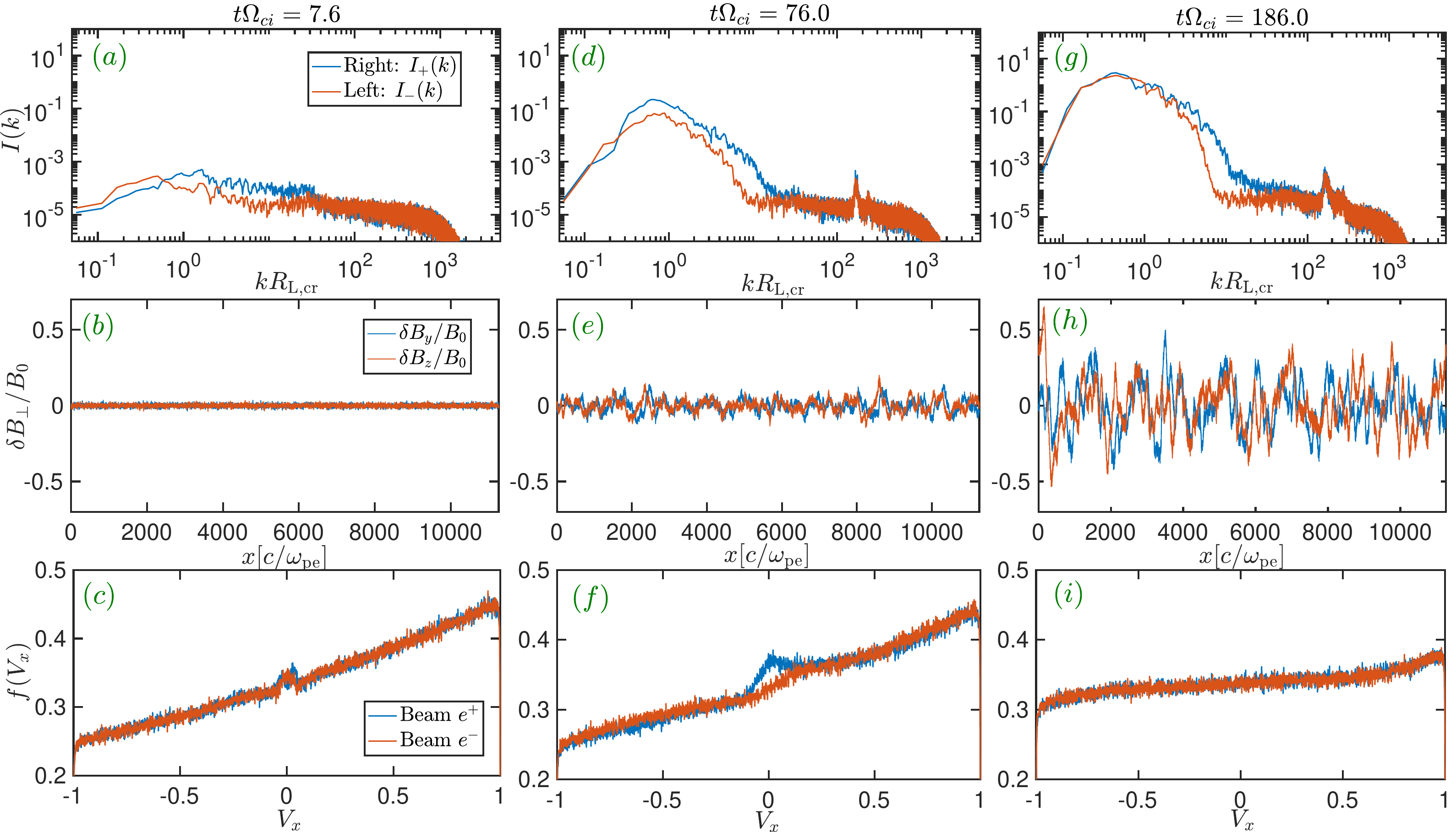}
	\caption{Wave spectrum $I(k)$ of right-handed and left-handed polarisation waves (top panels: \emph{a, d}, and \emph{g}), spatial profile of magnetic field fluctuations $\delta B_{y, z} / B_0$ (middle panels: \emph{b, e}, and \emph{h}) and distribution function of beam electrons and positrons $f(V_x)$ and different times of the simulation (bottom panels: \emph{c, f}, and \emph{i}). Left column corresponds to early time ($t\Omega_{\rm ci} = 7.6$), middle column to the linear phase ($t\Omega_{\rm ci} = 76$) and right column to the saturated phase of the instability ($t\Omega_{\rm ci} = 186$).} \label{fig:multipanel_snaps_Fid}
	\end{center}
\end{figure*}

More detailed insight into the evolution of the instability can be gained by considering the wave spectrum and the distribution function anisotropy of beam species at different characteristic times. In \autoref{fig:multipanel_snaps_Fid} we present the wave spectrum $I(k)$ (top panels), spatial profile of magnetic field fluctuations $\delta B_{y,z} / B_0$ (middle panels), and the anisotropy of the distribution function of beam populations $f(V_x)$ (bottom panels). From left to right, we present three representative times: early linear phase ($t\Omega_{\rm ci} = 7.6$), in the middle of the linear phase ($t\Omega_{\rm ci} = 76$), and the late post-saturation time ($t\Omega_{\rm ci} = 186$), respectively. The wave intensity in Fourier space for different polarisations $I_{\pm}(k)$ was done by transforming real-space magnetic fields $\delta B_{x,y}$ and according to $\int I(k)dk = \langle \delta B^2 / B_0^2\rangle $. Details of the method are presented in \autoref{sec:appendix_Fourier_construction}.

During the initial time of the evolution, at $t\Omega_{\rm ci} \sim 7.6$ (panels a-c) the wave spectrum is flat for both left-handed and right-handed wavemodes, the amplitude of magnetic fluctuations is negligible, and the distribution function of beam populations conserve its initial anisotropy.  $f(V_x)$ conserves its initial linear slope, except only slight change near the pitch angle $\mu = 0$ (i.e., near $V_x = 0$). In the middle of the linear phase of the instability, at $t\Omega_{\rm ci} = 76.0$, wave-modes with $0.1 \leq k R_{L,0} \leq 10$ are rapidly growing. The fastest growth occurs around $k R_{L,0} \approx 0.7$ as expected from the analytical calculation in \autoref{Eq:growth_rate_final}. Panel $e$ shows that the wave fluctuations reach $\delta B_{y,z} / B_0 \simeq 0.1$ level, and the distribution function of beam populations starts to be noticeably affected (panel $f$). We note that the asymmetry between left-handed and right-handed modes (blue and red lines in panel $d$, respectively) is mirrored by a different behaviour of beam positrons and beam electrons (blue and red lines in panel $f$, respectively). During the late time when the instability saturated (panels $g - i$) the wave spectrum is nearly stationary pertaining some asymmetry between left- and right-handed modes. There is a lack of right-handed modes in a narrow band of wavenumber $5 \leq k R_{L,0} \leq 20$. The amplitude of spatial fluctuations reached $\delta B / B_0 \simeq 0.3$, and the anisotropy in the distribution function of beam electrons and positrons is nearly erased. The latter is characterised by the quasi-flatness of $f(V_x)$.

An interesting feature is a line-like feature in the wave spectrum at short wavelengths at $k R_{L,0} \simeq 170$ or $k d_e \simeq 1.7$ (panels $d$ and $g$). While being energetically subdominant to the bulk of excited wavemodes at $k R_{L,0} \leq 10$ we, however, devote a subsection to the investigation of physical origin of this line feature at the end of this section (subsection \ref{subesect:line_feature}). 

\subsubsection{Linear phase}
\label{subsec:linear_PIC} 

\begin{figure}
	\begin{center}
	\includegraphics[width=0.93\columnwidth]{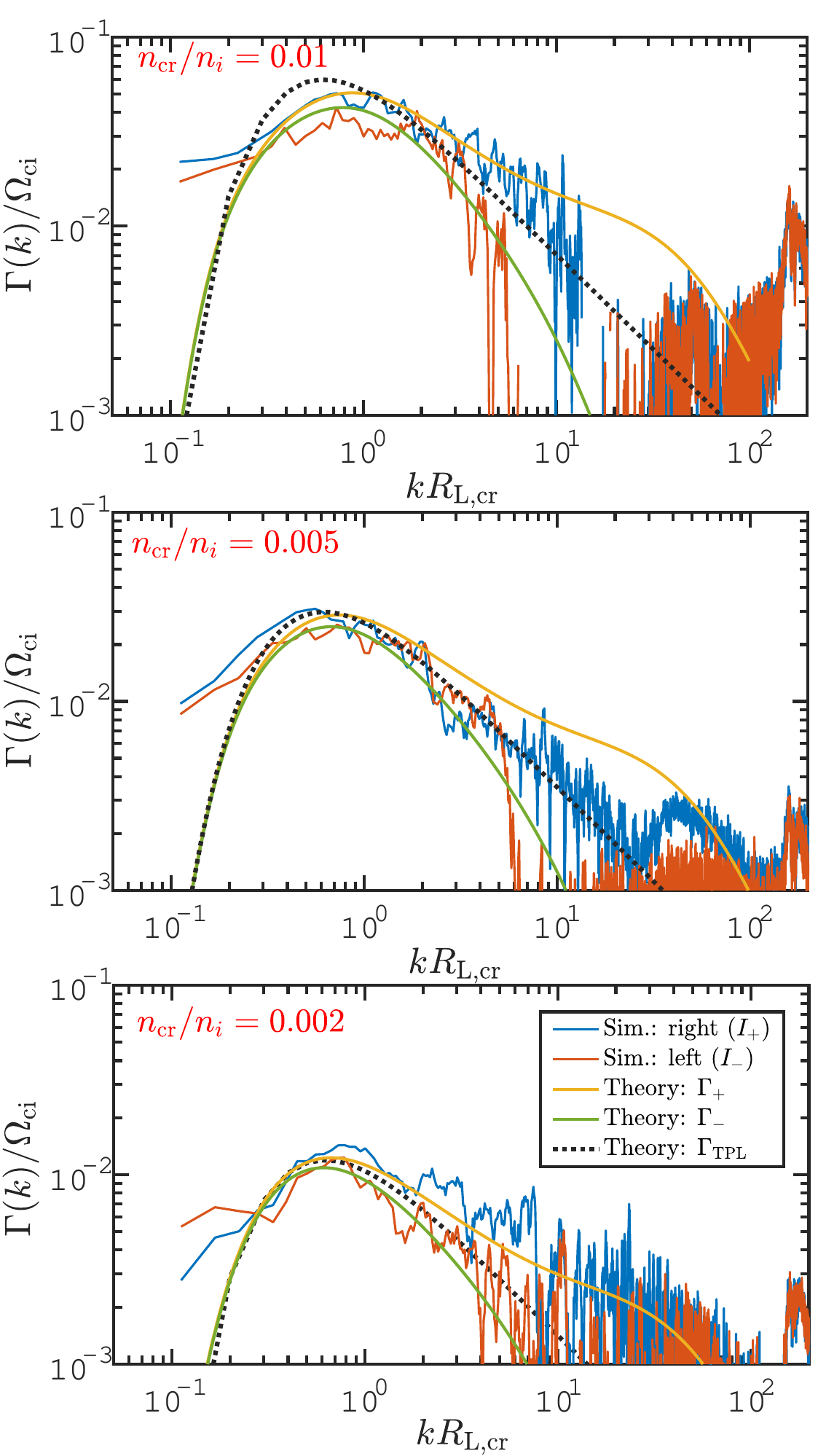}
	\caption{Linear growth rate of the instability $\Gamma$ as function of the wavenumber $k$ for two polarisation modes: circular right-handed (blue solid line) and left-handed (red solid line). Results for the simulations with three different values of CR density are presented: top panel corresponds to $n_{\rm cr} / n_i =0.01$, middle panel corresponds to $n_{\rm cr} / n_i = 5 \times 10^{-3}$, and bottom panel corresponds to $n_{\rm cr} / n_i = 2 \times 10^{-3}$. In each panel, the solid orange and green lines correspond to the theoretical growth rate using the solution of the full dispersion relation \autoref{Eq:HFDR}. Black doted line follows the expectation in test-particle limit from \autoref{Eq:growth_rate_final}.}\label{fig:growth_rate_k_by_k}
	\end{center}
\end{figure}

We now turn to more quantitative diagnostic of the linear phase of the instability. As expected, the accuracy of our numerical simulations (code and setup) can be tested versus the analytical expectations derived in the previous section. By performing spatial Fourier transform of the time-stacked magnetic field data $ FFT(B_{y,z} (x,t)) = \tilde{B}_{y,z}(k,t)$ we can follow the time evolution of intensity in different wave-modes $k$. We then select the approximate time when the growth is exponential ($\Delta t_{\rm lin}$) and then perform the linear fit of $\vert \log(\tilde{B}_{y,z}(k, \Delta t_{\rm lin})) \vert $. This procedure provides the numerical growth rate of different wavemodes, available in the simulation. The smallest wavemode is $k_{\rm min} = 2 \pi / L$ and the largest is $k_{\rm max} = \pi / \Delta$, where $\Delta$ is the spatial grid cell size \footnote{Waves at $k > \pi / \Delta$ are not resolved, as the Nyquist wavenumber is $2\pi / \Delta$.}. 

The result for three different simulations is presented in \autoref{fig:growth_rate_k_by_k}. In this figure we show the growth rates derived from simulations (solid blue and red lines), the basic analytical growth rate in the test-particle limit (black dotted line) and the growth rates with no low-frequency approximation (orange and green lines).  The blue-coloured line corresponds to the right-handed modes ($I_+$) while the red-coloured lines correspond to the left-handed modes (($I_-$). Different panels present results from simulations with different beam-to-background number density ratio: $n_{\rm cr} / n_i=0.01, 0.005, 0.002$ from top to bottom panel, respectively.

We can note that the maximum linear growth rate is obtained at $kR_{\rm L,cr} \simeq 0.7$ in agreement with analytical expectations from \autoref{Eq:growth_rate_final}, for all of three simulations. The growth rates are reasonably well reproduced up to $k R_{\rm L,cr} \sim 5-6$. At higher $k$ some differences appear in the different mode polarisation. Circular right-handed modes always produce higher growth rates than circular left-handed ones. This can be explained by adopting the high-frequency relation derived from \autoref{Eq:HFDR}. However, it is difficult to provide a firm conclusion because of the increasing effect of numerical noise with $k$. The agreement with analytical expectations is better for lower $n_{\rm cr} / n_i$, as can be seen by comparing the top panel with the bottom panel. This is expected since analytical calculations have been derived in the quasi-linear limit.

\subsubsection{Saturated phase}
\label{subsec:saturated_phase_PIC}

\begin{figure}
	\begin{center}
	\includegraphics[width=0.99\columnwidth]{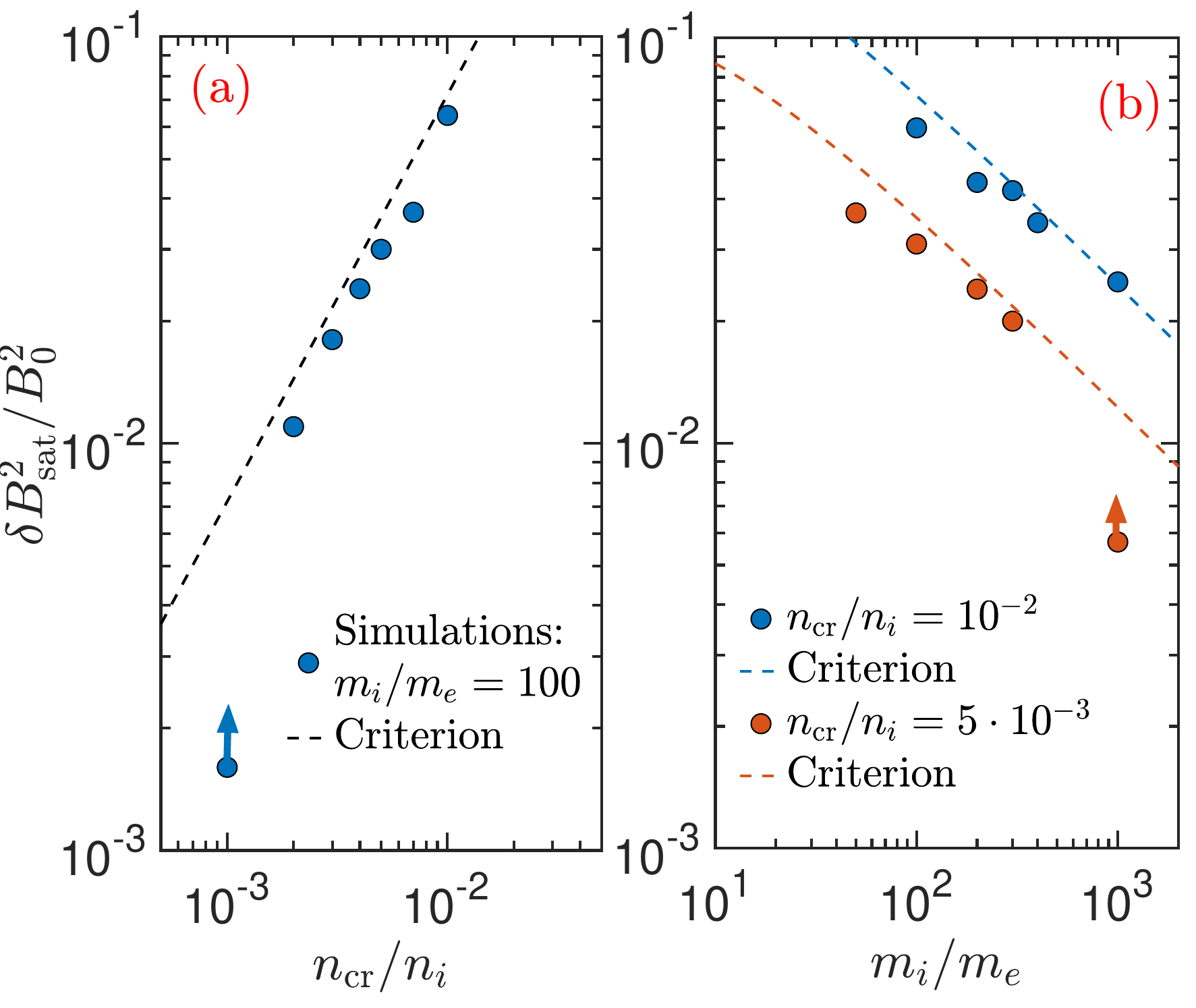}
	\caption{Saturation level of the magnetic fluctuations. \textit{Panel a}: Wave saturation intensity as function of $n_{\rm cr}/n_i$. Blue circles present the results from 1D3V PIC simulations. The red dashed line follows the scaling from \autoref{Eq:BSAT} that balances the momentum loss by CRs with momentum gained by waves. The data point at $n_{\rm cr} / n_i = 10^{-3}$ is a lower limit as the simulation ended before a saturated state was reached. \textit{Panel b}: same but varying $m_i / m_e$ at fixed $n_{\rm cr} / n_i = 10^{-2}$ (blue symbols) and $n_{\rm cr} / n_i = 5 \cdot 10^{-3}$ (red symbols).} \label{fig:deltaB_sat}
	\end{center}
\end{figure}

As previously discussed in Section~\ref{subsect:Overall_PIC} it can be identified that after the linear phase, where the wave growth is exponential, the wave-saturation occurs rapidly (see top panel in \autoref{fig:overall_evolution_Fid}). It corresponds to the phase when the wave intensity reaches its maximum and remains nearly constant afterwards. This level defines the saturation of wave intensity as $\delta B_{\rm sat}^2 = \delta B_{y, {\rm end}}^2 + \delta B_{z, {\rm end}}^2$.

In \autoref{fig:deltaB_sat} we show the dependence of saturation level of the magnetic field, $\delta B_{\rm sat}^2/B_0^2$, on the beam density $n_{\rm cr} / n_i$ (panel a) and on the mass ratio $m_i / m_e$ (panel b). The blue circle symbols present the results from different simulations where $n_{\rm cr} / n_i$ varies from $0.001$ to $0.01$; all other parameters being kept the same ($m_i / m_e =100$, $V_D = 0.1 c$, $V_A = 0.01 c$. See \autoref{table:PICparam}). The dashed black line follows the level derived from the criterion assuming that all of the available momentum lost by the decelerated CR beam (from $V_D$ to $V_A$) is transferred to the waves. This criterion was defined in \autoref{Eq:BSAT} \footnote{We used the fact that for an ultra-relativistic Maxwell-J\"{u}ttner distribution (see \autoref{Eq:MJ}) with $\theta = k_B T_{\rm cr} / m_e c^2 \gg 1$ the average particle momentum is $\langle p \rangle = 3 k_B T_{\rm cr} / c$}. The values measured in the simulations are slightly below the expectation but show very good agreement with the expected trend. We note that such a good agreement is remarkable for a PIC simulation and owing to the simplicity of the analytic criterion. \footnote{The exception is the simulation with the lowest $n_{cr} / n_i = 10^{-3}$. Here, the simulation was too short to reach saturation. We directly verified that the linear phase was still ongoing at the final time of the run.}

We also derived the saturation in simulations with different $m_i / m_e$ from $50$ to $1000$ (panel b of \autoref{fig:deltaB_sat}). The same criterion provides a good approximation of the saturation level for both $n_{cr} / n_i = 5 \times 10^{-3}$ and $n_{cr} / n_i = 10^{-2}$ cases. The `lack' of waves is small and can probably be imputed to other channels of dissipation (e.g., background plasma heating, some amount of wave damping).

It is interesting to note that the acquired final velocity of background $e^-$-ion plasma is roughly equal to the saturated wave intensity, $\langle V_i \rangle_{\rm bg} / V_A \simeq \delta B_{\rm sat}^2/B_0^2$. This was found in all simulations where both quantities could be measured with good confidence. The wave saturation criterion assumes that the momentum lost by CRs when decelerating from $V_D$ drift to $V_A$ is transferred to waves. Yet, the Alfv\'en waves are carried by the background fluid that seems to be dragged to the mentioned equality in the process of saturation of the instability.

\subsubsection{Properties of the wave spectrum}
\label{subesect:spectrum_asymetry_Kcut}

\begin{figure}
	\begin{center}
        \includegraphics[width=0.99\columnwidth]{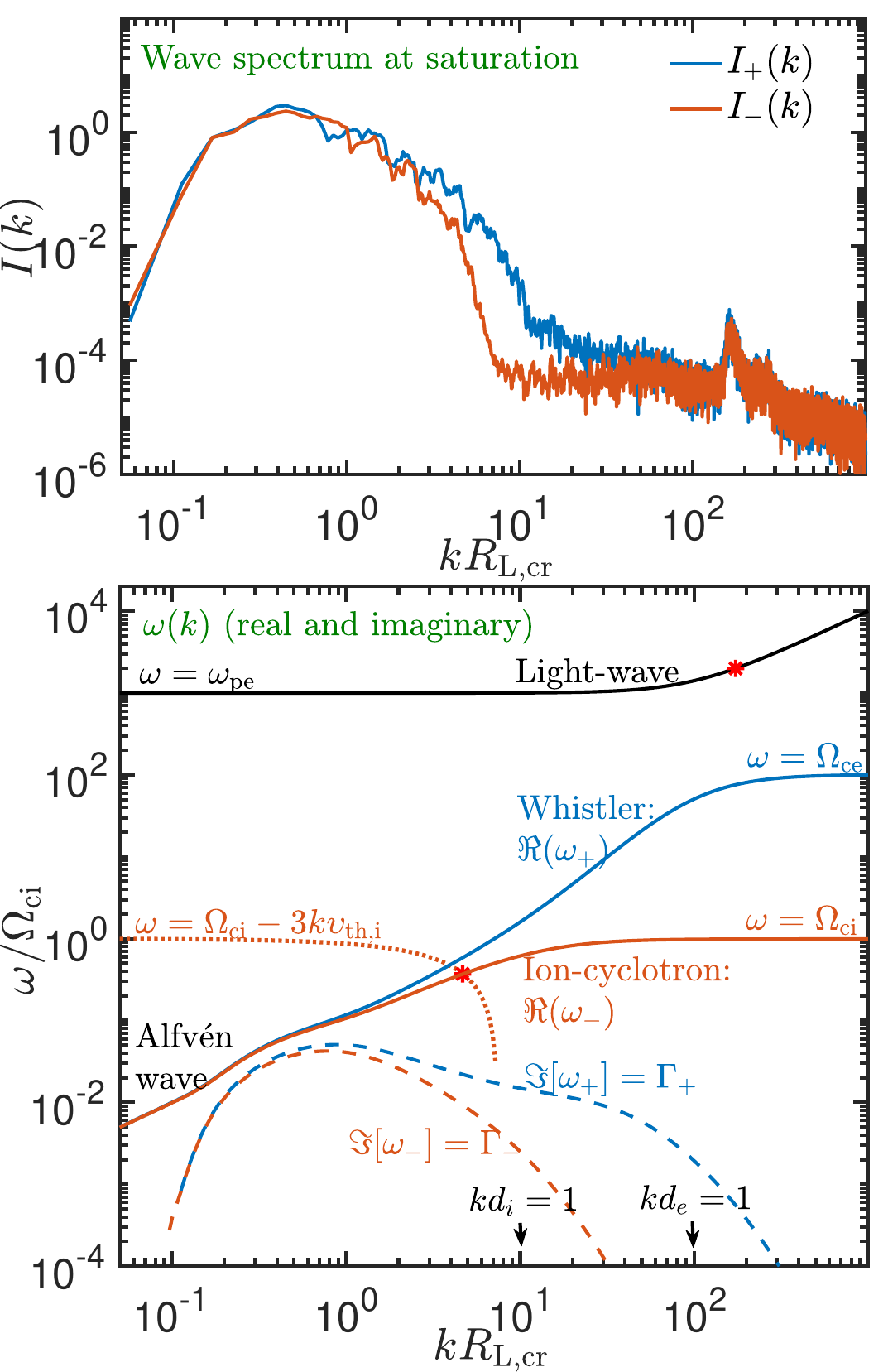}
	\caption{Wave spectrum at saturation and corresponding plasma wavemodes. \textit{Top panel}: wave spectra in Fid simulation. The blue line for right-handed waves and red line for left-handed waves. \textit{Bottom panel}: analytic dispersion relation of waves obtained by solving \autoref{Eq:HFDR}. The real part of $\omega$ (solid lines) and instability growth rate for the two polarizations that is the imaginary part of $\omega$ (dashed lines). The light-wave is shown with black solid line. The red star symbol at the top left side locates the intersection with $\omega = 2 \omega_{\rm pe}$ line. Alfv\'en wave corresponds to low frequency and low-$k$ part of red and blue solid lines. At $k R_{\rm L,cr} > 1$ the two branches separate into Whistler (blue solid line) and ion-cyclotron (red solid line). Vertical arrows in bottom panel locate the transition into ion-cyclotron regime (at $k d_i =1$) for left-handed waves, and electron-cyclotron regime $k d_e =1$ for right-handed waves. The lower red star symbol locates the intersection between the ion-cyclotron branch (red solid line) and $\Omega_{\rm ce} - 3k\upsilon_{\rm th,i}$ (red dotted line).} \label{fig:SatSpec_Dispersions}
	\end{center}
\end{figure}

Overall intensity is only one aspect of the self-generated waves. The properties of wave spectrum may require more subtle analysis. The wave spectrum after saturation in the Fid simulation, $I_{\rm sat} (k) = \delta B_{\rm sat}(k)^2 / B_0^2$, is shown in \autoref{fig:SatSpec_Dispersions}, top panel. The blue and red solid lines correspond to the final spectrum of right- and left-handed circularly polarised waves, respectively. Previously presented asymmetry between two polarisation modes is observed at $k R_{\rm L,cr} > 6$. A line-like feature is clearly seen at $k R_{\rm L,cr} \sim 170$.

In order to understand different cut-off positions for two modes we consider in more detail the proper wavemodes and theoretical growth rates. The corresponding result is presented in \autoref{fig:SatSpec_Dispersions}, bottom panel. The solutions of the full dispersion relation are obtained by solving \autoref{Eq:HFDR}. Both real and imaginary (unstable) parts are shown using solid and dashed lines, respectively. The light-wave is shown with a black solid line. Alfv\'en wave corresponds to low frequency and low-$k$ part of red and blue solid lines. At $k R_{\rm L,cr} \geq 4 - 6$ the two branches separate into Whistler (blue solid line) and Ion-cyclotron (red solid line). Alternatively, these two wave branches can be analytically derived with more ease by setting $\mathcal{C}=0$ (CR term) in the dispersion relation. Solving the equation without the CR term then provides accurate solutions for Whistler and Alfv\'en/ion-cyclotron (AIC) dispersion relations \citep[e.g.,][]{Vainio_2000}; which are identical to $\mathcal{R}$-labelled solutions shown in \autoref{fig:SatSpec_Dispersions} (bottom panel).
By comparing the dashed red/blue lines and solid red/blue lines, one can clearly see that the modes at $k R_{\rm L,cr} > 10$ are growing into the Whistler (right-handed) and ion-cyclotron (left-handed) wavemodes. This is the first hint to understand the asymmetry between $I_+$ and $I_-$ spectra.

\begin{figure}
	\begin{center}
        \includegraphics[width=0.95\columnwidth]{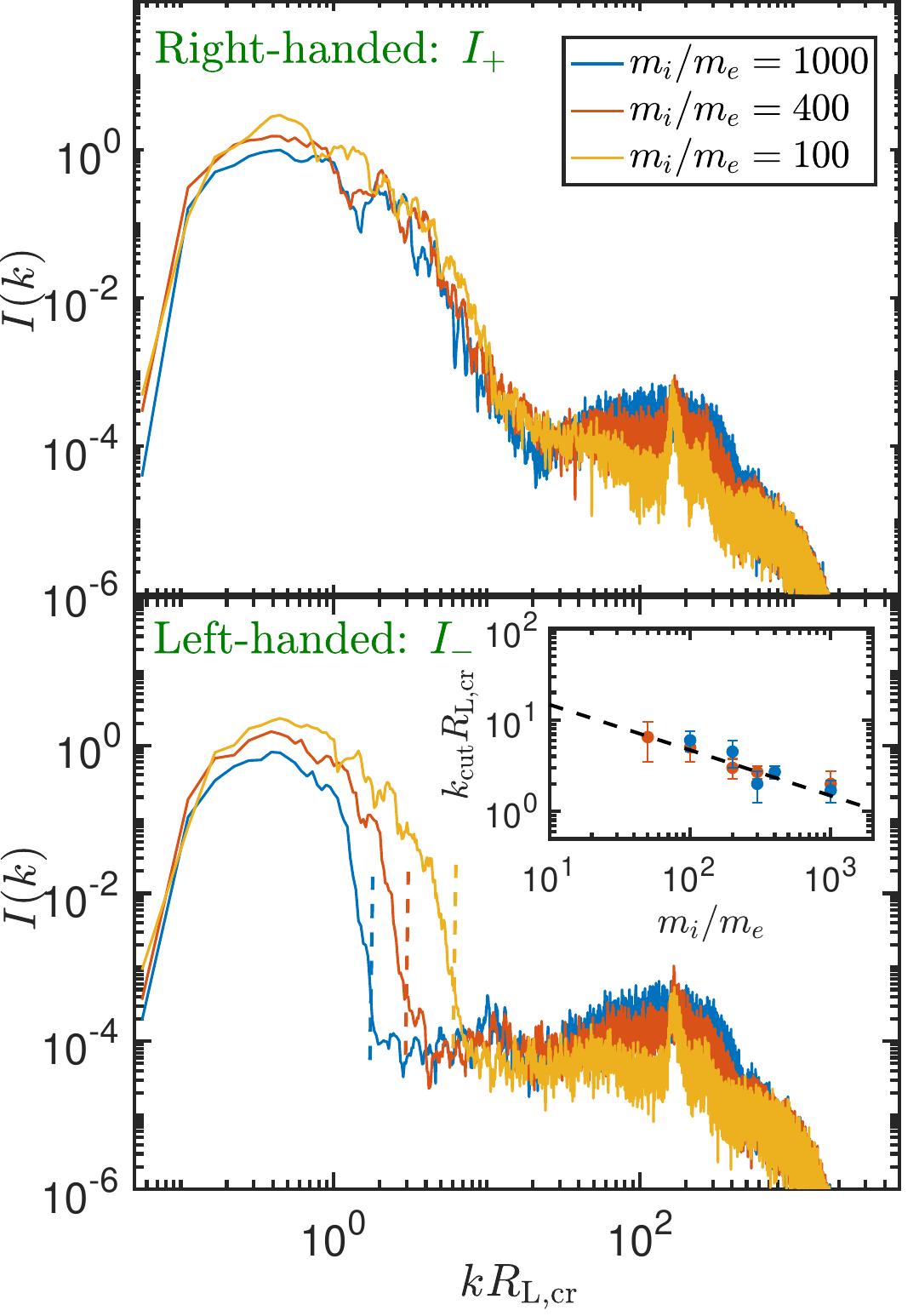}
	\caption{Saturated spectra of right-handed waves ($I_+$; top panel) and of left-handed waves ($I_-$; bottom panel). Lines of different colour correspond to different mass ratio adopted in the simulations: blue line corresponds to $m_i / m_e = 1000$ (Fid run), red line corresponds to $400$, and orange line corresponds to $100$. \textit{Inset:} cut-off wavenumber of left-handed branch for different values of $m_i / m_e$. Blue circles correspond to values derived from simulations with $n_{\rm cr} / n_i = 10^{-2}$ and red circles are from simulations with $n_{\rm cr} / n_i = 5 \times 10^{-3}$. The dashed black line follows $k_{\rm cut} d_i = 0.5$.} \label{fig:Kcut}
	\end{center}
\end{figure}

We now test the idea that the left-handed polarisation waves ($I_-$) are affected by ion-cyclotron resonance with background plasma. In \autoref{fig:Kcut} we present the saturated spectra of right-handed waves ($I_+$; top panel) and of left-handed waves ($I_-$; bottom panel). Lines of different colours correspond to different mass ratio adopted in the simulations: blue line corresponds to $m_i / m_e = 1000$ (Fid run), red line corresponds to $400$, and orange line corresponds to $100$. As can be seen in the top panel, $I_+$ spectrum is only weakly affected by the variation of $m_i / m_e$, while the $I_-$ branch (bottom panel) exhibits a cut that transits to lower $k R_{\rm L,cr}$ with increasing mass ratio. The position of the cut is highlighted using vertical dashed lines. For $m_i / m_e = 100$ the cut is located at $k R_{\rm L,cr} \simeq 6$, while for $m_i / m_e = 1000$ it is located at $k R_{\rm L,cr} \simeq 1.7$. 

More systematic analysis of $k_{\rm cut}$ value in large set of simulations is presented in the inset of the bottom panel. Blue circles correspond to values derived from simulations with $n_{\rm cr} / n_i = 10^{-2}$ and red circles are from simulations with $n_{\rm cr} / n_i = 5 \times 10^{-3}$. All other parameters are the same as in Fid run. 
The black dashed line follows $k_{\rm cut} d_i = 0.5$ scaling. We observe that all measured values cluster around this scaling. The basic interpretation is that all of $I_-$ waves are suppressed for any $k R_{\rm L,cr} > 0.5$ that falls in the regime where the Alfv\'en waves transition into Ion-cyclotron branch. 

An accurate location of resonant wavenumber can be found by considering the intersection between AIC branch (red solid line in bottom panel of \autoref{fig:SatSpec_Dispersions}) and $\omega = \Omega_{\rm ci} - 3 k \upsilon_{\rm th,i}$ (red dotted line in the same panel) \footnote{We found that the factor $3\upsilon_{\rm th,i}$ in this relation provides better agreement with simulation, instead of more standard $\upsilon_{\rm th,i}$. It was argued in \citet{Engelbrecht_2018} that this factor captures better the range of resonantly interacting background particles.}. This gives the cut-off at $k_{\rm cut} d_i = 0.47$, that is basically the same as the result plotted as the dashed line in the inset of \autoref{fig:Kcut}. In general, we found that the cut-off wavenumber agrees well with the scaling \citep{Engelbrecht_2018, Strauss_2019}:

\begin{equation}\label{Eq:Kcut_Strauss}
    k_{\rm cut,-} = \frac{\Omega_{\rm ci}}{V_A + 3 \upsilon_{\rm th,i}} \, .
\end{equation}

Left-handed waves with $k > k_{\rm cut, -}$ are expected to be strongly damped by background plasma. An improved calculation of the damping term following \citet{Strauss_2019} (Appendix B) shows that the resonance with first two ion-cyclotron harmonics suppresses all left-handed waves at $k > k_{\rm cut,-}$. Numerical solutions of the dispersion relation accounting for thermal effects in background plasma -- using plasma dispersion function -- shows as well that $k_{\rm cut,-}$ is an accurate estimate of both the measured values and of full solution of the dispersion relation. To convince ourselves that this is not a numerical effect, we note that the damping of ion-cyclotron waves is a well-observed process in solar wind plasma \citep[][for recent studies]{Woodham_2018, Bowen_2022}. The experimental cut-off scale agrees remarkably well with what we find in the present simulations.

Concerning the $I_+$ spectrum, the value of wavenumber ($k_{\rm cut,+}$) at which it drops to noise floor level should be regulated by the interaction with the Whistler branch. Using the physical parameters in Fid simulation, we found that the intersection of Whistler branch with $\omega = \Omega_{\rm ce} - 3 k \upsilon_{\rm th,e}$ occurs at $k d_e \simeq 0.56$. This value is in good agreement with \citet{Strauss_2019} (see their Figure~1 and Eqs. 61-64). However, Fid simulation shows that the cut-off of right-handed modes occurs at lower $k d_e \simeq 0.1$ (i.e., $k R_{\rm L,cr} \simeq 10$), see top panel of \autoref{fig:SatSpec_Dispersions}. This value is sensitively lower than expected $k_{\rm cut,+}$. By inspecting in more detail, we found that there are indeed growing waves up to $k d_e \sim 0.3 - 0.5$ earlier in the simulation. However, the part of spectrum at $k d_e > 0.1$ is damped during the saturated phase. Alternatively, in the simulations with lower CR number density ($n_{\rm cr} / n_i \leq 5 \times 10^{-3}$) the waves at $0.1 < k d_e < 0.5$ emerge more from the background level and remain at relatively low-intensity level. We have no satisfactory explanation for this different damping effect that increases for larger $n_{\rm cr} / n_i$. Yet, no wave growth is observed for $k d_e > 0.5$ in any simulation, that is consistent with expected $k_{\rm cut,+}$ value.

As a side check, we performed a simulation where the background plasma is composed of pairs ($m_i = m_e$, run PIC-17 in \autoref{table:PICparam}). The detailed presentation is reported in \autoref{sec:appendix_pairs_PIC}. To summarise, we found that there is no asymmetry between two polarisation modes of self-excited waves. The initial drift of CRs (beam population) reduced from initial $V_D = 14.1\,V_A$ to exactly $V_A$ for both CR electrons and positrons in an identical way. This confirms the idea that the asymmetry is due to the ionic nature of background plasma.

\subsubsection{Line feature in the wave spectrum}
\label{subesect:line_feature}

\begin{figure}
	\begin{center}
        \includegraphics[width=\columnwidth]{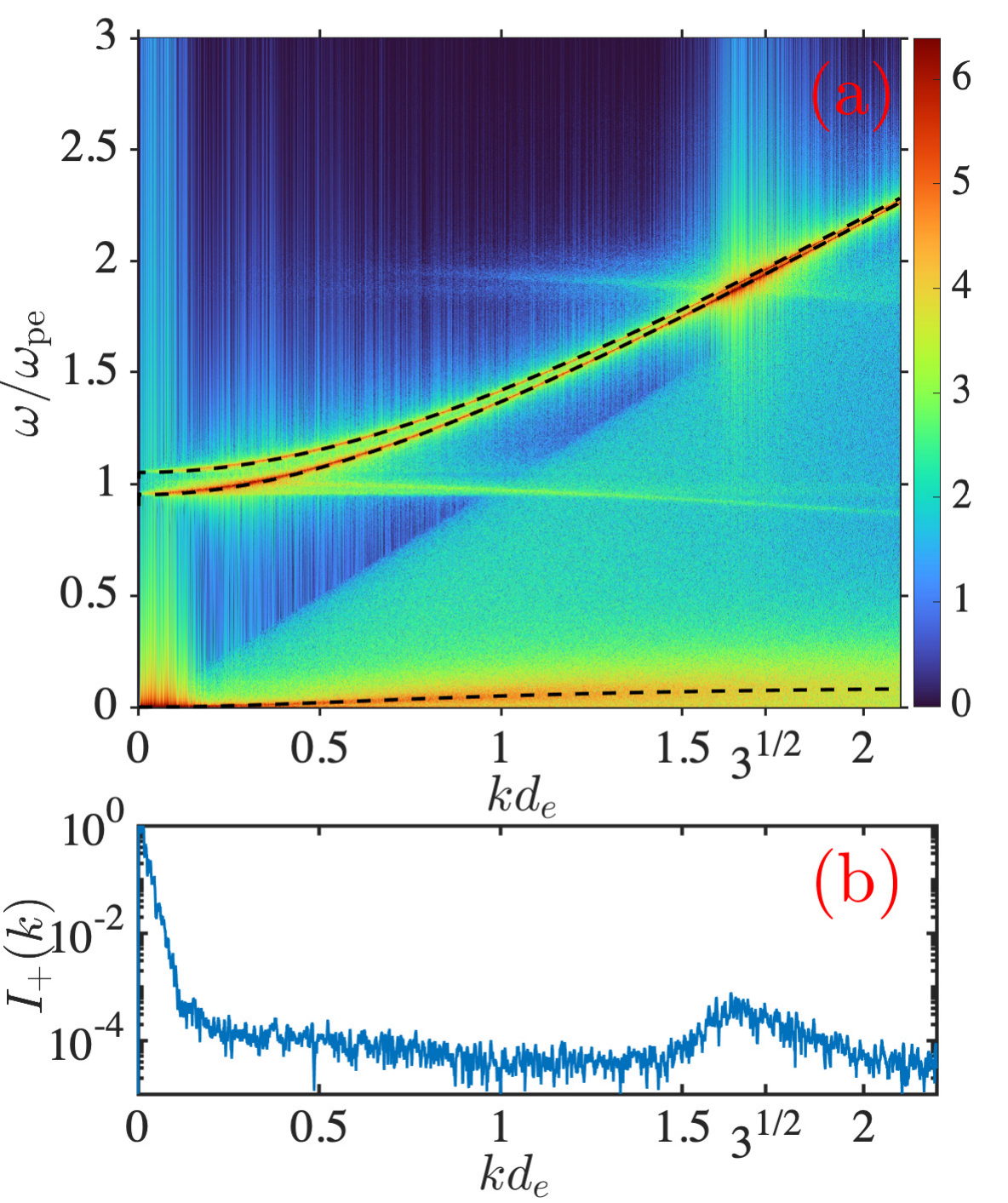}
	\caption{Wave dispersion analysis. {\it Top panel (a)}: Wave dispersion $\omega(k) / \omega_{\rm pe}$ at the final times of the linear phase. From Fid simulation (see \autoref{table:PICparam} for parameters). The diagnostic was performed on the $\delta B_z$ component between $t \Omega_{\rm ci} = 95$ ant $t \Omega_{\rm ci} = 96.6$  (at the end of the linear phase of the instability). Electromagnetic light-wave branch with left and right-handed polarisation is easily identifiable. At low $\omega < \Omega_{\rm ce}$ the Whistler branch is clearly seen as well (theoretical dispersion line over plotted with dashed line). Plasma emission at $\omega = \omega_{\rm pe}$ is observed and the second harmonic can be observed as well, especially in the brighter zone where it intersects the EM wave. \textit{Bottom panel (b)}: $I_+(k)$ spectrum at saturation from Fid simulation. It is the same spectrum as in \autoref{fig:SatSpec_Dispersions} but plotted using linear $x-$axis scale and wavenumber plotted in units of $d_e^{-1}$. This improves the visibility of the high-$k$ part of the spectrum.} \label{fig:dispersion_omega_k}
	\end{center}
\end{figure}

A minor effect of the interaction between CR beam and background plasma produces a high-$k$ line-like feature (e.g., feature at $k R_{\rm L,cr} \simeq 170 $ in \autoref{fig:multipanel_snaps_Fid}). It was observed in the wave spectra of all simulations except when background plasma is composed of pairs (see \autoref{sec:appendix_pairs_PIC}). As the spectra are only in $k$ space we possess limited information on the frequency of this wave. In order to clarify the nature of this line, we performed dispersion analysis to derive $\omega(k)$ from the simulation. This was done by storing the EM fields on the spatial grid at every time-step between $t \Omega_{\rm ci} = 95$ ant $t \Omega_{\rm ci} = 96.6$. It corresponds to the end of the linear phase of the instability. By stacking the snapshots of the stored fields and performing 2D FFT on a given field component we derive the dispersion relation $\omega(k)$.

In \autoref{fig:dispersion_omega_k} we present the $\omega(k)$ dispersion spectrum performed on $B_z$ component of the magnetic field. The light-wave (EM wave) is well-seen with dispersion $\omega^2 \simeq k^2c^2 + \omega_{\rm pe}^2$. As the plasma is magnetised, at low-$k$ the EM wave separates into right- and left-handed circular polarisation branches that have slightly different cut-off frequencies. The right-handed mode has the cut-off frequency at $\omega_{\rm RH} \simeq \omega_{\rm pe} + \Omega_{\rm ce}/2$ while the left-handed branch has the cut-off frequency at $\omega_{\rm LH} \simeq \omega_{\rm pe} - \Omega_{\rm ce}/2$ \citep{Kulsrud_2005}. Exact -- textbook -- solutions for both branches are over-plotted using black dashed lines. It can be seen that there is a region with enhanced intensity on this branch. This corresponds to a resonance between the EM wave and second harmonic $\omega  = 2\omega_{\rm pe}$. In terms of the wavenumber, theoretically, it corresponds to $k d_e = \sqrt{3}$. That is in good agreement with the measured position of the instantaneous spectrum $I(k)$, shown in the (b) panel of the figure. This spectrum is the same as presented in \autoref{fig:multipanel_snaps_Fid} and \autoref{fig:SatSpec_Dispersions} but using linear $x$-axis scale and the wavenumber normalized to $d_e^{-1}$ instead of previously used $R_{\rm L,cr}^{-1}$. In this panel, one can identify the dominant contribution of $k d_e < 0.1$ waves and a much less intense region slightly below $k d_e = \sqrt{3}$. Both panels combined lead to a firm conclusion that this line feature is the second-harmonic emission. We verified in a large set of simulations that the position of this line is always at $k d_e \simeq \sqrt{3}$, independently of $n_{\rm} / n_i$, $m_i / m_e$, and $T_{\rm cr}$ values explored.

There is extensive literature on PIC simulations of second-harmonic plasma emission in unmagnetized plasma \citep[e.g.,][]{Kasuba_2001, Henri_2019, Krafft_2021} and magnetized plasma \citep[e.g.,][]{Lee_2022}. In the present work, we do not focus on this interaction and limit on basic observation. This is also justified by the fact that this line-emission is largely subdominant to the main instability generated waves in Alfv\'en branch. It does not seem to have an important effect on the dynamics of the pair beam-driven instability and, more generally, on particle scattering properties.

For completeness, let us describe other noticeable features in \autoref{fig:dispersion_omega_k}. At low $\omega < \Omega_{\rm ce}$ the Whistler branch is clearly seen as well (theoretical dispersion line over plotted with dashed line). The dynamically important part (and by far the brightest spot) of the spectrum corresponds to the region $\omega / \omega_{\rm pe} < 0.1$ and $k d_e < 0.1$. It is not resolved in this figure as the time window for diagnostics was too small.

\section{Discussion}\label{sect:DIS}

This article addressed the pair-driven instability when the beam propagates through ionic background plasma at super Alfv{\'e}nic speed. We used fully kinetic particle-in-cell simulations in order to understand the non-linear evolution of the interaction.  Combined with detailed analytical derivation in appropriate approximation we verified the accuracy of linear phase of the instability and confirmed some intuitive features, such as a decrease of the beam velocity to near-Alfv\'enic speed.
The intensity of waves at saturation was found to be in good agreement with the momentum transfer criterion: momentum lost by beam particles when they decelerate (erase the anisotropy in the distribution function) is transferred to Alfv\'en waves. Both Alfv\'en and kinetic scales were covered by the wavemodes triggered by the instability. This leads to important damping of waves belonging to the ion-cyclotron and Whistler branches. All short-scale waves with $k d_i \gg 1$ are strongly suppressed by ion-cyclotron resonances with background plasma. We also observed a second harmonic plasma emission. These results might be of interest as input for detailed modelling of particle propagation in TeV halos.

Let us note, however, that by construction our analysis is limited to parallel waves as our simulations were 1D in space. There is no contribution from $k_\perp$ related phenomena. Yet, it is known that at kinetic scales with $k_{\perp} d_i >1$ there is a dominant contribution of kinetic Alfv\'en waves (KAW), that is well-studied in multidimensional kinetic simulations \citep[e.g., ][]{Cerri_16, Cerri_17}. A study of the interplay between self-generated parallel waves by the streaming instability and external turbulence is a topic for future work \citep[recently, a theoretical framework to address this problem was proposed by][]{Cerri_24}.

\subsection{Simulations rescaling in comparison with realistic halos}\label{S:SCAL}

Here, we discuss the relevance of our work to TeV halos, limitations, and possible extensions using other numerical techniques. To start, let us examine how realistic the parameter set chosen in our simulations is. We consider the case of the Geminga pulsar to provide a rough estimate of the ratio $n_{\rm cr}/n_i$ in a realistic halo. 

The calculation is based on the parameters adopted in \citet{2023PhRvD.107l3020S}, namely a typical pulsar age of $342$\,kyrs, a present spin-down luminosity of $L_{\rm age} =3.26 \times 10^{34}$ erg/s and an initial spin-down timescale $t_0 \simeq 12$\,kyrs. Under the assumption of pulsar spin down from magnetic dipole radiation (breaking index of 3), the luminosity at time $t$ is
\begin{equation}
\label{Eq:psrlum}
    L(t) = L_{\rm age} \frac{(1+ t_{\rm age}/t_0)^2}{(1+t/t_0)^2} \, .
\end{equation}
This implies an initial spin-down luminosity $L_0 =2.9 \times 10^{37}$ erg/s at early times $t \ll t_0$, three orders of magnitude above the present-day value. The particle injection spectrum follows a broken power law\footnote{The injection spectrum is also expected to feature a time-varying cutoff at an energy $E_c$, for instance to reflect the progressive decrease of the maximum potential drop, which would lead to $E_c(t) = 1.7 \sqrt{L(t) / 10^{36}~\rm{erg/s}}$\;PeV. We neglect this here because the high typical values for this cutoff energy have no impact on our particle density estimates.}
\begin{eqnarray}
\label{Eq:injspec}
    Q(E,t) &= Q_0(t) \left({E \over E_b}\right)^{-a_L}~\quad \rm{for}~E\leq E_b \nonumber \\
              &= Q_0(t)  \left({E \over E_b}\right)^{-a_H}~\quad \rm{for}~E> E_b \nonumber \ ,
\end{eqnarray}
where we use $E_b \sim 200$\,GeV, $a_L =1.6$, $a_H=2.4$, in agreement with the ranges of values considered by \citet{2023PhRvD.107l3020S}. 
The particle population is energized by the conversion of a constant fraction $\eta \simeq 1$ of the pulsar spin-down power, such that the normalisation $Q_0$ of the injection spectrum is obtained by 
\begin{equation}
\label{eq:normspec}
    \int_{E_{\rm min}}^\infty dE E Q(E,t) = \eta L(t), \eta \le 1 \ .
\end{equation}
The minimum particle energy is taken as $E_{\rm min} = 1$\,GeV. 
The average particle energy in the injection spectrum can be approximated as
\begin{align}
\label{Eq:eavg}
   \langle E \rangle &= H(a_L,a_H,r) \times E_b \\
   &\simeq \frac{(2-a_L)^{-1} + (a_H-2)^{-1} -(2-a_L)^{-1} r^{2-a_L}}{(a_H-1)^{-1} - (a_L-1)^{-1} + (a_L-1)^{-1} r^{1-a_L}} \times E_b \, , \nonumber
\end{align}
where $r=E_{\rm min}/E_b$. For the adopted parameters, this yields $H \simeq 0.12$ hence $\langle E \rangle \simeq 24$\,GeV. If $E_{min} = 100$ MeV then $H \simeq 0.03$ and $H$ goes as $r^{0.6}$ as $r \rightarrow 0$. The factor $H$ is weakly sensitive to $a_L$ and $a_H$, for instance keeping $E_{\rm min}=1$ GeV with $a_L=1.8$ (1.4) and $a_H=2.2$ (2.6) we find $H \simeq 0.1$ (0.17).

We assume that, at each time $t$, the particle injection rate $Q(E,t)$ is directly released across the surface of the nebula $S_{PWN}$. The present-day nebula can be approximated as a sphere with radius $R_{\rm pwn} = 0.1$\,pc \citep{2017ApJ...835...66P}, but it was likely much larger in the past, when the pulsar was more powerful. The typical size of a bow-shock nebula is the so-called stand-off distance, where pressure in the spherically expanding pulsar wind equals the ram pressure exerted by the ambient medium on the moving pulsar \citep[see Eq. 16 in][]{Gaensler:2006}. This stand-off distance scales as $\sqrt{L}$ and so $R_{\rm pwn}(t) = R_{\rm pwn}(t_{\rm age})\sqrt{L(t) / L_{\rm age}}$ and $S_{PWN} \propto L$.

The number density of pairs injected by the nebula at time $t$ can therefore be written as
\begin{equation}
\label{Eq:injdens}
   n_{\rm cr}(t) = \frac{ \eta L(t)  }{ \langle E \rangle S_{PWN} V_D } \, .
\end{equation}
For a meaningful comparison with the situation implemented in the simulations, we assume that the particles stream across the surface at a velocity $V_D = 10 V_A$, where $V_A$ is the Alfv\'en speed in the surrounding medium. This numerical choice seems reasonable since $V_D$ is basically a free parameter. The typical background gas density around Geminga can be estimated from the absence of $H_\alpha$ emission surrounding the Geminga bow shock to be in the range between $n_i \sim 10^{-3}~\rm{cm^{-3}}$ \citep{2003Sci...301.1345C} and $n_i \sim 10^{-2}~\rm{cm^{-3}}$ \citep{2017ApJ...835...66P}. We use the latter higher bound together with a magnetic field strength of $1\mu$G to compute $V_A = 22$ km/s.

According to \autoref{Eq:injdens}, the particle density immediately beyond the surface of the nebula does not depend on time because $S_{PWN} \propto L$ and all other parameters are considered constant. Using present day values for $L =3.26 \times 10^{34}$\,erg/s, $R_{\rm pwn} = 0.1$\,pc and other parameters being specified above (i.e., $\eta = 1$, $V_D = 10 V_A = 220$\,km/s, $\langle E \rangle = 24$\,GeV), we get $n_{\rm cr} \sim 3\times10^{-8}~\rm{cm^{-3}}$. Such a value may be considered as a lower limit because the above calculation assumes instantaneous transfer of the pulsar power to the escaping particle flux, whereas there is most likely an accumulation of particles within the nebula for centuries or millennia or more before they can escape or lose their energy. On the other hand, the values of $n_i$ and $B_0$ adopted above favour a small Alfv\'en speed in the surrounding medium, hence a small escape velocity $V_D$ and therefore a higher particle density. Overall, the beam-to-background density ratio $n_{\rm cr} / n_i$ that can be expected close to the injection zone, that is close to the pulsar wind nebula, is likely in the range $10^{-6}-10^{-5}$ (with the caveat that the background density at early times might have been higher than $10^{-2}~\rm{cm^{-3}}$). The PIC simulations presented here are therefore not representative of the conditions in the Geminga system.

Another important difference with the astrophysical case resides in our choice of particle distribution. In our simulations, we consider a Maxwell-J\"{u}ttner distribution and most of the energy is carried by particles around $10$\,MeV instead of $200$ GeV. Hence, a realistic simulation should have a beam-to-gas density ratio in the range $10^{-6}-10^{-5}$ and a beam with a bulk energy of $E_b \sim 10^5 m_e c^2$. It seems that if the resonant streaming instability can be triggered in the environment of evolved pulsars it should be in a region close to the pulsar wind nebula, otherwise the surface term in Eq. \ref{Eq:injdens} produces too much beam dilution. Actually, these values require an unrealistically large numerical box as the PIC method has to resolve the electron skin depth (ideally Debye length). Such high energies would require the use of larger-scale simulations, including a magnetohydrodynamic (MHD) description of the plasma. We therefore discuss briefly in the next subsection an alternative approach based on combining PIC and MHD but leave a more complete study of this regime for future work.

We note that we did not provide a direct extrapolation of the results from our PIC experiments to astrophysical scales for the following reasons: (i) the convergence of simulations with low values of $n_{\rm cr} / n_i \leq 10^{-3}$ was not reached, (ii) the saturation level of waves confirms intuitive scaling that is already used in the literature in general, and (iii) employing specific non-periodic boundary conditions is needed for direct comparison with a given astrophysical system.

\subsection{Large-scale simulations: MHD-PIC}
As discussed above, the main limitation is the scale separation that we are able to achieve with full-PIC simulation. This was reflected by the adopted distribution of CRs with an average energy of $E \sim 10$\,MeV.
Using MHD-PIC method \citep{2015ApJ...809...55B2015ApJ...809...55B, vanmarleetal:2018, Mignone_2018}, we can reach higher energies and lower beam densities, which reflect more closely the type of beam-plasma interaction found in astrophysical circumstances. To demonstrate the possibilities of the MHD-PIC method, we performed two simulations in a 2-D box measuring $50\times0.5\,R_L$, with $R_L$ the Larmor radius of the peak of the beam energy distribution. This box is subdivided into a grid of 2000 by 20 grid cells. All boundaries are set to periodic.
The box is filled with a stationary thermal plasma and a magnetic field aligned with the $x$-axis in which the Alfv{\'e}n speed is set at $0.01\,c$, the plasma being in equipartition. Through this plasma, we run a beam of non-thermal particles with an advection speed of $0.1\,c$, and a thermal particle distribution according to a Maxwell-J{\"u}ttner prescription with the partial motion being isotropic in the co/moving frame. This beam is made up of an average of 50 particles per particle species per grid cell. 
We run two MHD-PIC simulations. One with a temperature of $T_{\rm cr} = 10\,m_ec^2$ and a relative particle density of $n_{\rm cr} / n_i = 5\times10^{-3}$ for comparison with the PIC simulations. The second simulation has a beam temperature of $T_{\rm cr} = 100\,m_ec^2$ and a relative particle density of $n_{\rm cr} / n_i = 10^{-3}$ to demonstrate MHD-PIC's ability to reach a more realistic astrophysical parameter space.  The simulation is allowed to run while the gas and particles relax. 

\begin{figure}
	\begin{center}
        \includegraphics[width=0.9\columnwidth]{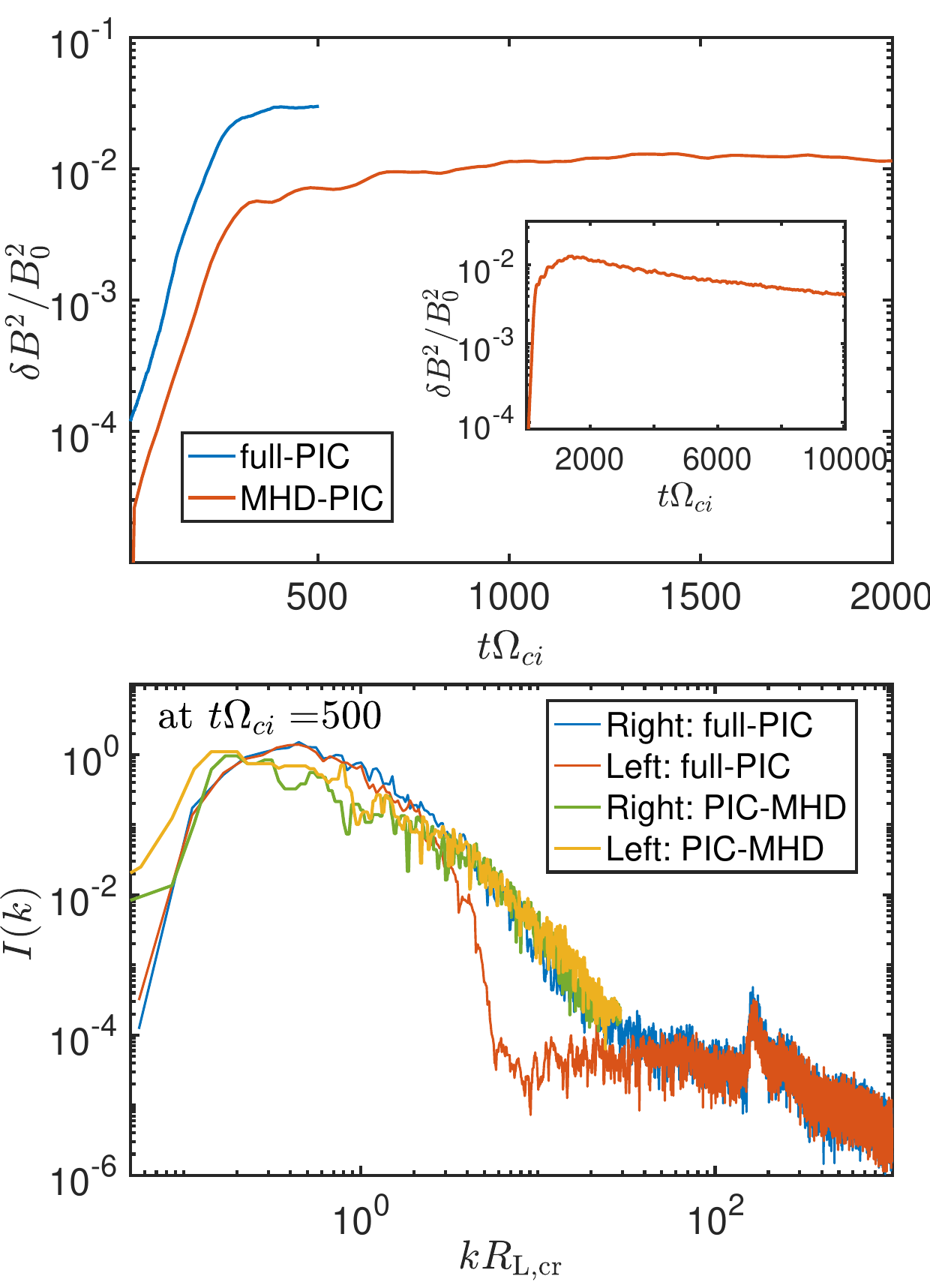}
	\caption{Comparison of the results from full-PIC and MHD-PIC simulations using identical parameters as in run \emph{PIC-3} given in \autoref{table:PICparam}. The top panel presents the average magnetic energy evolution in time for full-PIC (blue solid line) and MHD-PIC (red solid line) techniques. The inset shows the full time-evolution from MHD-PIC simulation, as it runs over a much longer timescale than any full-PIC run. The bottom panel shows the comparison of wave spectra at the time that corresponds to the final time of the full-PIC simulation. Blue and red solid lines show the spectra from full-PIC for right- and left-handed modes, respectively. Green and orange solid lines show the same for MHD-PIC simulation. \emph{Note:} the spectra from MHD-PIC simulation were upshifted by a factor of $\approx 2$ in order to get a better visual comparison.} 
 \label{fig:compare_PIC_MHDPIC}
	\end{center}
\end{figure}

For the simulation with $T_{\rm cr} = 10\,m_ec^2$ and $n_{\rm cr} / n_i = 5\times10^{-3}$ (same parameters as full-PIC simulation \emph{PIC-3}). The result of this comparison is presented in \autoref{fig:compare_PIC_MHDPIC}. As can be seen in the upper panel of the figure (magnetic energy evolution), during the linear phase the slope in both PIC and MHD-PIC is very close. Meaning that the fastest growth rates are in good agreement with both techniques. We found that the peak magnetic wave intensity value was lower in MHD-PIC by a factor of $\approx 2$ (difference in peak values for blue and red lines). Yet, the wave spectra (bottom panel) were comparing reasonably well, except for the absence of the cut in wave spectra of left-handed modes at $k d_i >0.6$. This is expected, as the resonant ion-cyclotron absorption effect is absent if the background plasma is treated using an MHD formalism.

\begin{figure}
	\begin{center}
        \includegraphics[width=0.99\columnwidth]{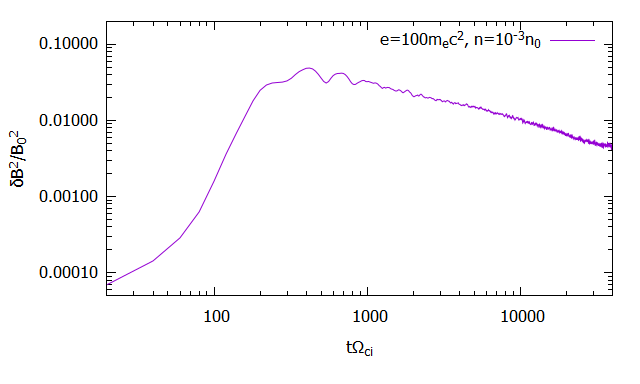}
	\caption{Magnetic energy evolution of a MHD-PIC simulation with $T_b\,=\,100m_ec^2$ and relative beam density $n_{\rm cr} / n_i = 10^{-3}$.} 
 \label{fig:highElowD}
	\end{center}
\end{figure}

For the simulations with a temperature of  $100\,m_ec^2$, the resulting magnetic wave energy evolution in our simulation is shown in Fig.~\ref{fig:highElowD}. The maximum value $\delta B_{\rm sat}^2 / B_0^2 \sim 0.05$ is reached at $t \Omega_{\rm ci} \approx 400$. The level of magnetic fluctuations at the saturation state is consistent with the expected value from \autoref{Eq:BSAT} which gives $\delta B_{\rm sat}^2 / B_0^2 \sim 0.05$ for the adopted parameters (we also used $m_e/m_i = 100$). 
From Fig.~\ref{fig:highElowD}, we can also estimate the growth rate of the magnetic field instability for the linear phase, which lasts from $t\Omega_{\rm ci}~\simeq~90$ until $t\Omega_{\rm ci}~\simeq~180$. From our simulation, we find a growth rate over that period of approx. $\Gamma/\Omega_{ci} \sim 3\times10^{-3}$ in reasonable agreement with the expected value from Eq. \ref{Eq:growth_rate_final}.

This technique is much less computationally prohibitive than full-PIC simulations and could potentially lead to regimes which are closer to astrophysical parameters. Some improvements such as $\delta f$ method could be implemented to reach better signal-to-noise ratio and distribution function representation \citep[e.g.,][]{Denton_1995, Bai19} as well as non-periodic boundary conditions \citep{Bai_2022} that are important to properly describe the saturation of the instability. This work is deferred to a forthcoming study.

\section{Conclusion}

To conclude, we investigated in detail the evolution of a resonant electron-positron beam-driven streaming instability that develops in a cold background electron-ion plasma. We considered a pair beam following a Maxwell-J\"uttner distribution and drifting with respect to the background plasma at a super-Alfvénic speed with initial value  $V_D = 10 V_A$. The simulations were performed using 1D3V Particle-in-Cell simulations. \\

We find the following main results:
\begin{enumerate}
    \item Maximum linear growth rates at scales $kR_{\rm L,cr} \sim 0.7$ are well recovered. The solutions at smaller scales show polarisation-dependent growth rates, with forward right modes taking over left-handed modes. This effect can be explained by relaxing the condition over the wave frequency in the moving frame to be $\ll \Omega_{\rm ci}$. At high $k$ the right and left-handed modes follow the Whistler/electron-cyclotron and the Alf\'en/ion-cyclotron branches respectively. The simulations show a reasonable agreement with the analytical solutions. At $k \gg d_i^{-1}$ left-handed modes are damped due to ion-cyclotron resonance and the right-handed modes are cut due to the interaction with the Whistler branch.

    \item We have checked that the criterion for magnetic saturation by momentum transfer from the beam to the waves is well reproduced, with the exception of the low beam density case where the simulations have not converged.
\end{enumerate}
Our results underline the importance of kinetic simulations when subtle effects determine the saturation level and properties of the self-generated spectrum.

We also compared PIC simulations with 1D3V MHD-PIC simulations using the same setup. In particular, we find a reasonable agreement between the two methods. MHD-PIC results show a bit more dissipation, leading to a typical saturated magnetic field energy that is a factor $\approx 2$ below the one obtained using PIC simulations (see \autoref{fig:compare_PIC_MHDPIC}). 

We finally discuss the rescaling of our numerical setup to the context of gamma-ray halos around evolved pulsars. We find that the kinetic numerical setups have beam-to-background density ratios still several orders of magnitude above the one currently expected in the Geminga pulsar halo. Moreover, the non-thermal particle population carrying the bulk of momentum has typical kinetic energies several orders of magnitude above the range covered by PIC simulations and is unreachable for now. MHD-PIC simulations may partly fill the energy gap to describe the gamma-ray halo more properly. This deserves a future investigation.

\begin{acknowledgements}
The authors are grateful to C. Evoli and B. Olmi for fruitful discussions. C. Evoli also provided us with an estimation of the beam-to-background plasma density ratio, which was helpful for the discussion in section \ref{S:SCAL}. I.P is grateful to Silvio Sergio Cerri and Gabriel Fruit for insightful suggestions that improved our understanding of this topic. We acknowledge financial support by ANR for support to the GAMALO project under reference ANR-19-CE31-0014. The simulations have been performed on CALMIP supercomputing resources at Universit\'e de Toulouse III (France) under the allocation P20028.
\end{acknowledgements}


\bibliographystyle{mnras}
\bibliography{main_biblio} 


\begin{appendix}
\section{Electron-positron background plasma} \label{sec:appendix_pairs_PIC}

  \begin{figure}
	\begin{center}
	\includegraphics[width=0.95\columnwidth]{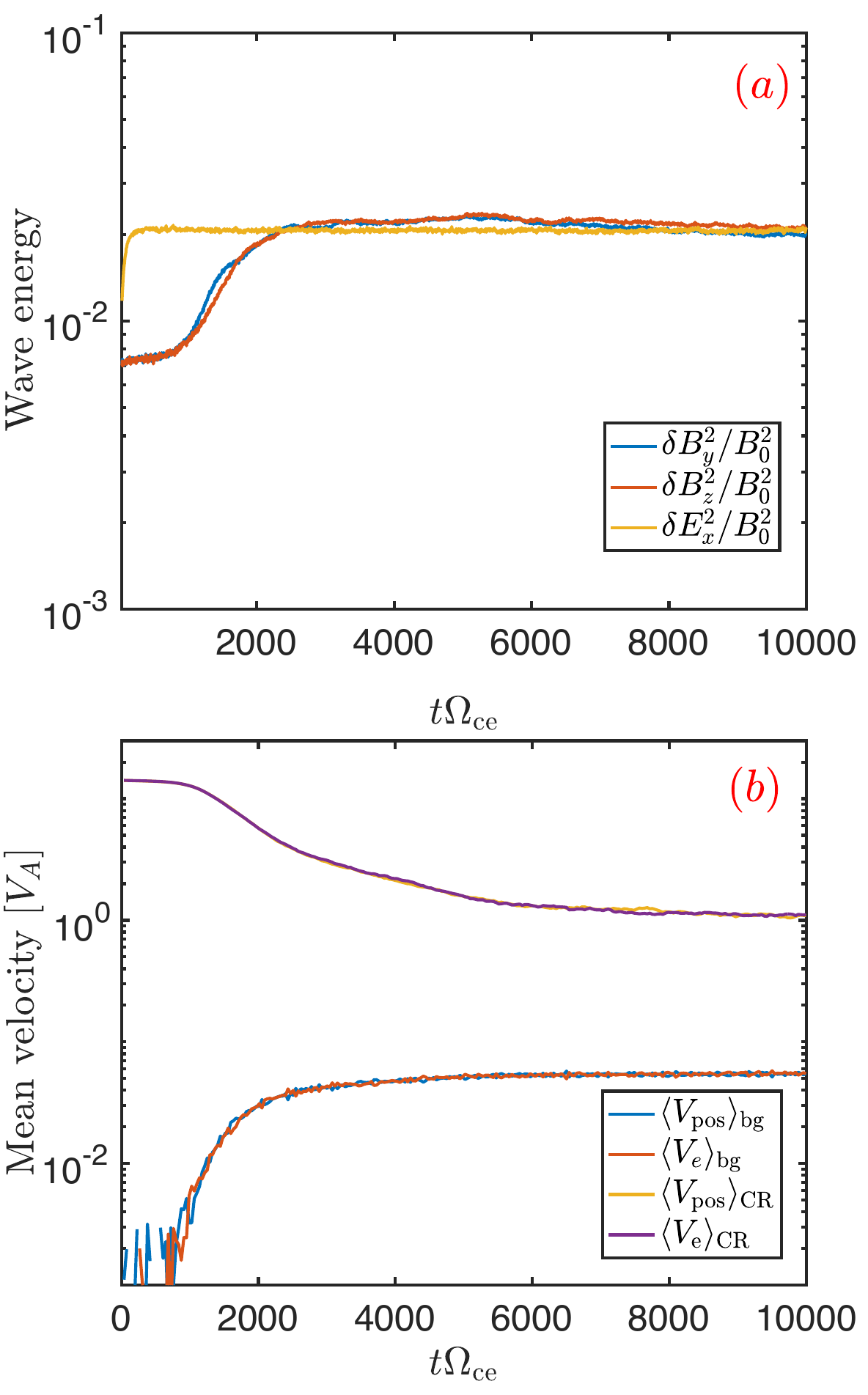}
	\caption{Same as \autoref{fig:overall_evolution_Fid} but for the simulation with electron-positron background $m_i = m_e$; run PIC-17. Overall time evolution of field intensities (panel a) and drift velocities of different populations (panel b).} 
	\label{fig:overall_evolution_pairs}
	\end{center}
\end{figure}

We want to test whether the asymmetry in the dynamic behaviour of CR positron population with respect to the CR electron population  (see Figure 1) and if the spectrum asymmetry is due to the ionic nature of the background plasma. In this appendix, we simplify the simulation setup by considering background plasma being composed of $e^{\pm}$ pairs ($m_i = m_e$) instead of ion-electron case studied in the main text. Indeed, such a setup imposes mass and charge symmetry that minimises space charge effects.

Due to typical scales change and the fact that the background mass is no longer dominated by ions we also changed several other parameters as follows: $m_i = m_e$, $n_{\rm cr} / n_i = 10^{-3}$, $T_{\rm cr} = m_e c^2$; $\Omega_{\rm ce} / \omega_{\rm pe} = 0.01$. The initial drift was unchanged $V_D / c = 0.1$, and background temperature $T_{\rm bg} = 10^{-3} m_e c^2$. The deduced Alfv\'en speed is $V_A = \Omega_{\rm ci} / \omega_{\rm p} =  \Omega_{\rm ce} / \omega_{\rm p} = 10^{-2} / \sqrt{2}$, where the division by $\sqrt{2}$ is due to $m_i = m_e$, hence total plasma frequency is $\omega_p = \sqrt{2} \omega_{\rm pe}$. These parameters are also reported in \autoref{table:PICparam}, simulation PIC-17.
This setup ensures that the energy density in CRs (kinetic + rest mass) is not dominant compared to the background.

In \autoref{fig:overall_evolution_pairs} we present the time evolution of wave energy averaged over the simulation box and drift velocities of four populations (background electrons and positrons, and CR electrons and positrons) \footnote{This figure is basically the same as \autoref{fig:overall_evolution_Fid} but for the $e^{\pm}$ background simulation.}. Despite a high level of noise floor measured in the longitudinal electric field $E_x$, we can clearly identify the linear phase of the instability between $t\Omega_{\rm ce} \simeq 1000$ and $t\Omega_{\rm ce} = 1600$. The saturation occurs after $t\Omega_{\rm ce} = 2500$. The intensity of self-generated waves settles at $\delta B_{\rm sat}^2 / B_0^2 \approx 0.45$. 
As seen in the bottom panel, the decrease in CR drift is identical for CR electrons and CR positrons. Both drop at the same rate from $\sim 14 V_A$ to $V_A$, with good accuracy. Concerning the background plasma, it gains some amount of momentum corresponding to final drift $\langle V_{\rm bg} \rangle_{\rm pos} = \langle V_{\rm bg} \rangle_e \simeq 0.055 V_A$.

\begin{figure*}
	\begin{center}
	\includegraphics[width=0.99\textwidth]{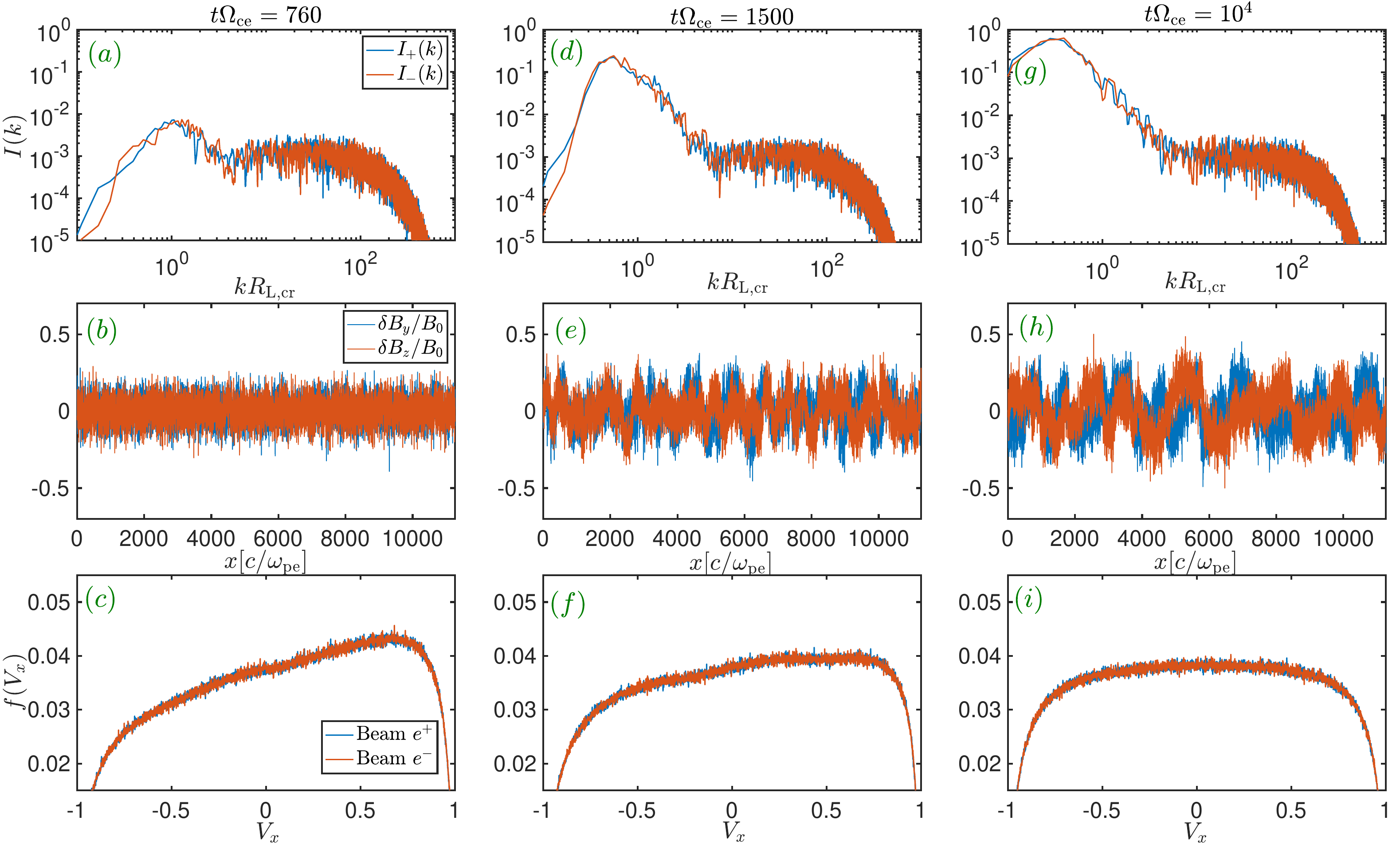}
	\caption{Wave spectrum $I(k)$ (top panels), magnetic field spatial profile (middle panels) and distribution function of beam electrons and positrons $f(V_x)$ and different times of the simulation. Left column corresponds to early time, middle column to the linear phase and right column to the saturated phase of the instability.} \label{fig:multipanel_snaps_pairs}
	\end{center}
\end{figure*}

Concerning the wave spectrum and anisotropy in CR distribution function, in \autoref{fig:multipanel_snaps_pairs} they are presented in the same format as in \autoref{fig:multipanel_snaps_Fid} of the main text. Both branches of wave polarisation are identical at any time of the simulation. Their intensity at three different times is presented with blue and red lines in the top row of the figure. The same is true for the anisotropy in the distribution function (bottom row of the figure). Both CR electrons and positrons undergo a reduction of anisotropy at the same rate. It also reflects in identical behaviour of overall drift velocities. The final state presents a roughly flat $f(V_x)$, such as the anisotropy being basically erased.
We also note that the line-like feature in the high-$k$ region of the wave spectrum, observed in simulations with $m_i \gg m_e$, is absent here.

  \begin{figure}
	\begin{center}
	\includegraphics[width=0.9\columnwidth]{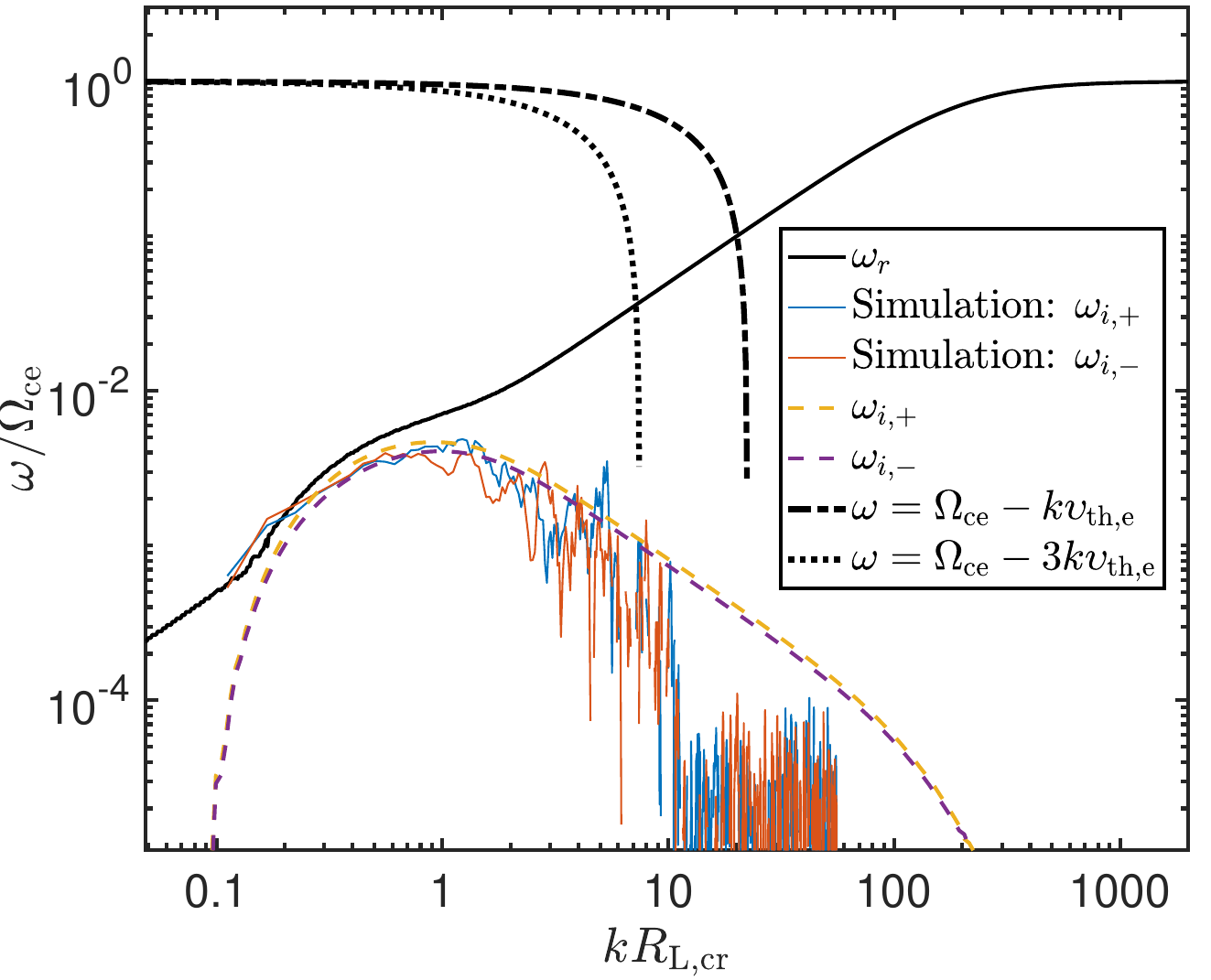}
	\caption{Real (black solid line) and imaginary part (dashed lines) of the dispersion relation obtained by solving \autoref{Eq:HFDR}, in the case where the background is pair plasma. The growth rates from the corresponding simulation are plotted using blue and red solid lines for right- and left-handed modes), respectively. The fit was not performed on $k R_{\rm L,cr}>50$ modes as being strongly affected by thermal noise. The black dot-dashed and dotted lines correspond to Doppler-shifted cyclotron resonance of background thermal plasma:  $\omega = \Omega_{\rm ce} - k \upsilon_{\rm th,e}$ and $\omega = \Omega_{\rm ce} - 3 k \upsilon_{\rm th,e}$, respectively.} 
	\label{fig:dispers_Grate_pairs}
	\end{center}
\end{figure}

To complete the analysis of this simulation, we report the derived linear growth rate of the instability in \autoref{fig:dispers_Grate_pairs}. The agreement between simulation and analytical calculation is good up to $k R_{\rm L,cr} \simeq 10$. As in the Fid simulation, we had to use the full dispersion equation that does not assume $\omega \ll \Omega_{\rm ce}$. The test-particle limit solution deviates clearly from the measured rate, especially around the fastest-growing wavenumber. We note that waves are not growing at $k R_{\rm L,cr}>10$. We recall that the theoretical growth rates were derived in approximation of cold background plasma. But when background thermal effects are considered, the Doppler-shifted cyclotron resonance with thermal electrons and positrons removes waves with $k R_{\rm L,cr}>20$ ($k R_{\rm L,cr}>7$) if the intersection between $\omega_r$ and $\omega = \Omega_{\rm ce} - k \upsilon_{\rm th,e}$ ($\omega = \Omega_{\rm ce} - 3 k \upsilon_{\rm th,e}$) is considered. In \autoref{fig:dispers_Grate_pairs} this effect can be seen as an intersection between the black solid line ($\omega_r$) and black dot-dashed line (dotted line, respectively). This explains the absence of self-generated waves in $k R_{\rm L,cr}>7$ region observed in \autoref{fig:multipanel_snaps_pairs}. At saturation, it seems that the resonance condition with factor of 3, $\omega = \Omega_{\rm ce} - 3 k \upsilon_{\rm th,e}$, provides better agreement for the measured cut-off location in the simulation, consistent with what was found in the main text for the cut-off of $I_{-}(k)$, see Section \ref{subesect:spectrum_asymetry_Kcut}.

We conclude that the different dynamical behaviour of CR electrons and CR positrons, the asymmetry in the self-generated wave spectrum and line-like emission feature are absent if the background is $e^\pm$ plasma. All these features require the background to be of ionic nature.

\section{Linear growth rates: mono-energetic and power-law cases}\label{sec:appendixgrowth}

For completeness, we provide below the growth rates in the quasi-linear limit for power-law and mono-energetic distributions. These two distributions can physically occur in realistic astrophysical systems and could be explored in future numerical simulations. Expressions given below could be useful as analytic expectations for such study.

\subsection{Mono-energetic distribution}
In the case the beam has a mono-energetic distribution we have $g(p)=p_0^{-2} \delta(p-p_0)$. It comes \begin{equation}
    \chi_{\rm cr} = i \left(1- {k V_D \over \omega}\right) {2\pi^2 e^2 \over k \omega} {n_{\rm cr} \over p_0}  \ .
    \label{eq:chi_CR-M}
\end{equation}
\subsection{Power-law distribution}
In the case the beam has a power-law distribution we have for the dimensionless function $g(p)=(\alpha-3) p_0^{-3} \left(p/p_0\right)^{-\alpha}$ \citep{2018PhRvD..98f3017E}, where $p_0 \le p \le p_{\rm max}$. It comes if $\alpha > 2$:
\begin{equation}
    \chi_{\rm cr} \simeq i\left(1- {k V_D \over \omega}\right) {2\pi^2 e^2 \over k \omega} {n_{\rm cr} \over p_0} {(\alpha-3) \over (\alpha-2)} \left({p_{\rm min}(k)\over p_0}\right)^{2-\alpha} \ .
    \label{eq:chi_CR-P}
\end{equation}

\section{Wave decomposition in Fourier space} \label{sec:appendix_Fourier_construction}

Here we describe the procedure of construction of the wave intensity in Fourier space, $I(k)$. We follow the same procedure as presented in \citet{Bai19} (Appendix A), but without assuming that the waves are strictly Alfvénic. The main lines are described below.

The wave intensity $I(k)$ is constructed according to 
\begin{equation}
    \int_{k_{\rm min}}^{k_{\rm max}} I(k) dk = \left \langle \frac{\delta B^2}{B_0^2} \right \rangle \ .
    \label{eq:Ik_normalisation}
\end{equation}

The brackets $\langle \cdot \rangle$ stand for spatial average.
A way to decompose circularly polarised modes is to consider the following combination:
\begin{equation}
C_\pm(x)=\frac{\delta B_y\pm{\rm i}\delta B_z}{B_0}\ .
\end{equation}
A discrete Fourier transform of $C^\pm(x)$,
\begin{equation}
W_\pm(k_i)=\frac{1}{N}\sum_{n=0}^{N-1}C^\pm(x_n)e^{-{\rm i}k_ix_n}\ ,
\end{equation}
gives the wave amplitude and phase. The wavenumber of mode $i$ is defined as $k_i=\frac{2\pi}{L_x}i$ and $L_x = N \Delta$ is the length of the simulation box. The wave intensity is then given by
\begin{equation}
I_\pm(k)=|W_\pm(k)|^2\frac{L_x}{2\pi}\ .
\end{equation}
With this construction, one defines the dimensionless power in modes of wavenumber $k$ per logarithmic scale as $\mathcal{F}(k) = k I(k)$, since $dk = 2\pi / L_x$. For clarity, we mainly used $I(k)$ in the figures of the main part of the article.

\end{appendix}
\end{document}